\documentclass[12pt]{article}
\usepackage{amsfonts,amsmath,amssymb}
\usepackage{graphicx}

\newfont{\twelvemsb}{msbm10 scaled\magstep1}
\newfont{\eightmsb}{msbm8}
\newfam\msbfam
\textfont\msbfam=\twelvemsb \scriptfont\msbfam=\eightmsb
\catcode`\@=11
\def\Bbb{\ifmmode\let\next\Bbb@\else
\def\next{\errmessage{Use \string\Bbb\space only in math mode}}\fi\next}
\def\Bbb@#1{{\fam\msbfam{{#1}}}}

\newcommand{\be}{\begin{equation}}
\newcommand{\ee}{\end{equation}}
\newcommand{\ba}{\begin{eqnarray}}
\newcommand{\ea}{\end{eqnarray}}

\begin{document}
\sloppy
\renewcommand{\thefootnote}{\fnsymbol{footnote}}

\newpage
\setcounter{page}{1} \vspace{0.7cm}
\begin{flushright}
LAPTH - 1138/06\\
PTA/06 - 1 \\
March 2006
\end{flushright}
\vspace*{1cm}
\begin{center}
{\large \bf On the finite size corrections of anti-ferromagnetic anomalous dimensions in ${\cal N}=4$ SYM}\\
\vspace{1.8cm} {\large G.\ Feverati $^a$, D.\ Fioravanti $^b$, P.
\ Grinza $^c$ and M.\ Rossi $^a$ \footnote{E-mail addresses:
feverati@lapp.in2p3.fr, fioravanti@bo.infn.it,
grinza@LPTA.univ-montp2.fr, rossi@lapp.in2p3.fr.}}\\
\vspace{.5cm} $^a${\em Lapth \footnote{UMR 5108 du CNRS,
associ\'ee \`a
    l'Universit\'e de Savoie.}, 9 Chemin de Bellevue, BP 110, F-74941 Annecy-le-Vieux Cedex, France} \\
\vspace{.3cm} $^b${\em Department of Mathematics, University of York (UK),\\
Department of Physics and Mathematics, University of Trento (Italy),\\
INFN and Department of Physics, University of Bologna (Italy)} \footnote{The last one from December 19-th 2005 onward.}\\
\vspace{.3cm} $^c${\em LPTA, Universit\'e Montpellier II, Place
Eug\`ene Bataillon, 34095 Montpellier, France}
\end{center}
\renewcommand{\thefootnote}{\arabic{footnote}}
\setcounter{footnote}{0}
\begin{abstract}
{\noindent Non-linear} integral equations derived from Bethe
Ansatz are used to evaluate finite size corrections to the highest
(i.e. {\it anti-ferromagnetic}) and immediately lower anomalous
dimensions of scalar operators in ${\cal N}=4$ SYM. In specific,
multi-loop corrections are computed in the $SU(2)$ operator
subspace, whereas in the general $SO(6)$ case only one loop
calculations have been finalised. In these cases, the leading
finite size corrections are given by means of explicit formul\ae
\, and compared with the exact numerical evaluation. In addition,
the method here proposed is quite general and especially suitable
for numerical evaluations.
\end{abstract}
\vspace{1cm} {\noindent {\it Keywords}}: Bethe Ansatz;
integrability; non-linear
integral equation; spin chains; super Yang-Mills theories; AdS/CFT correspondence. \\

\newpage

\section{Prologue}
\setcounter{equation}{0}

The AdS/CFT correspondence relates string states and local
gauge-invariant operators of a dual quantum field theory
\cite{MWGKP}. The energies of the string states correspond to the
eigenvalues (the so-called anomalous dimensions) of the mixing
matrix of gauge field theory operators. However, even for the well
understood case of the ${\cal N}=4$ super Yang-Mills (SYM) theory,
testing of such a correspondence has revealed to be a rather
difficult task. Indeed perturbative expansions in SYM assume the
't Hooft coupling $\lambda $ to be small, while string
perturbation theory makes sense when the string tension $T={\sqrt
{\lambda}}/{2\pi}$ is large. A conceptual progress came with the
proposal \cite {BMN} of restricting to states (named afterwards
BMN states) represented by small closed strings with large angular
momentum $L$. Indeed, the energy of these states admits an
expansion in the small classical parameter $\lambda/L^2$, the
quantum corrections corresponding to another, $1/L$ expansion. On
the SYM side, these string states correspond to composite
operators containing $L$ local fields. However, this implies that
the (for now) accessible semiclassical regime of string states is
mimicked in the gauge theory by ``long'' operators, which renders
a priori the computation of their anomalous dimensions a complex
problem. In this perspective, it can be understood why a very
important progress was realised by Minahan and Zarembo while
noticing the coincidence between the one-loop expansion of the
mixing matrix for operators in the ${\cal N}=4$ SYM theory
containing $L$ local fields and an integrable $SO(6)$, $L$ sites
spin chain Hamiltonian \cite{MZ}. For, since then, many Bethe
Ansatz ideas and techniques were used in order to find anomalous
dimensions of very long operators of gauge theories, as long as
this novel {\it correspondence} between Integrable Models (IMs)
and (S)YM theories was extended to the full $PSU(2,2|4)$ symmetry
\cite {BS}, to higher loop expansions \cite{BKS, SS, BDS} and to
other gauge theories \cite{WW,Roi, Bere, FHZ,DVT}. For
completeness' sake, we should also mention how this IM/SYM
relation makes more complete and deeper the previous appearance of
an integrable system in the barely perturbative regime of QCD
\cite{Lip, FadKo}. In particular this link to IMs stimulated an
impressive activity, which allowed many scholars to test the
AdS/CFT duality in different cases (e.g. for the BMN operators of
\cite{BMN} and for others too). More precisely, the integrability
of the mixing matrix at all orders in perturbation theory was
conjectured in \cite{BKS, BDS} and then proved for the $SU(2)$
subsector up to three loops in \cite{SS}. In this paper, the
dilatation operator was embedded into a long-range spin chain, the
Inozemtsev spin chain \cite{I}. However, the Inozemtsev spin chain
at four loops would lose the apparently desirable property (in
perturbative gauge theory) of BMN scaling and this lack stimulated
the conjecture according to which an alternative long-range spin
chain for all number of loops may exist \cite{BDS}. A Bethe Ansatz
was also proposed in \cite{BDS} for all the values of the coupling
$g$ (cf. below), in order to deal with this otherwise unknown
multi-loop Hamiltonian. However, the intriguing recent paper
\cite{RSS} has pointed out many reasons why this Bethe Ansatz may
not furnish the right anomalous dimension at and beyond the
wrapping order $g^{2L}$ (note also \cite{AJK} for a qualitative
interpretation). The same paper has identified in the half filled
Hubbard model a short range Hamiltonian conjectured to reproduce
the mixing matrix. Actually, this identification was explicitly
carried out up to three loops and it is unclear if it may survive
the break-down order $g^{2L}$.

Despite the great amount of work on the subject, the majority of
the results found up to now (with exception of very few papers
like, for instance, \cite{HLPS,BTS,GK}) concerns the calculation
of only the leading term of the anomalous dimensions of arrays of
$L$ operators in the limit $L \rightarrow \infty$. In the BMN
sector, this corresponds to classical energies on the string side.
Consequently, the correction to this leading term is related to
quantum fluctuations of the string state energy and is therefore
worth studying. In this respect and from the Bethe Ansatz point of
view the $L \rightarrow \infty$ limit may be described in all its
physical quantities in terms of the density of Bethe roots (per
quantum numbers), provided the latter really tends to a continuous
distribution when $L=\infty$ (cf. \cite{Fad81} for some remarks on
this point). In any event, almost as early as the Bethe's
invention \cite{Bethe} (for the spectrum of the isotropic
Heisenberg spin $1/2$ chain), a linear integral equation
constraining the density (for the anti-ferromagnetic ground state)
was derived and solved \cite{Hulten}. Since those early stages the
power and versatility of Bethe Ansatz was being very much
appreciated, at most in condensed matter physics, integrable
models theory~\footnote{The first liaison with the quantum version
of integrability, the Yang-Baxter equation, was found by the
seminal contributions \cite{Yang-Baxter}.} and statistical
mechanics (cf. \cite{Fad95, KBI} just for some examples). Also,
the integral equation idea lived a revival since 1964
\cite{Mcguire, Yang-Baxter} and was extended to excited states
(cf. the profundus review \cite{deVega}) and to the statistical
view \cite{TakSuz}. Implementing this latter in the framework of
the relativistic factorised scattering theory, Al. Zamolodchikov
formulated a general and pretty widely applicable idea concerning
an exact formula of the vacuum scaling function at all size
scales, the so-called thermodynamic Bethe Ansatz \cite{ZamTBA}.

As for finite size effects in quantum integrable systems, the
Non-Linear Integral Equation (NLIE) description -- first
introduced in \cite{KBP} for the conformal (anti-ferromagnetic)
vacuum and then derived for an off-critical vacuum in \cite{DDV}
by other means -- turned out to be an efficient tool in order to
explore the scaling properties of the energy. Since \cite{FMQR},
regarding excitations on the vacuum, a number of articles was
devoted to the analysis of and through a NLIE and mainly follows
the route pioneered by Destri and de Vega \cite{DDV} (cf. the
lectures \cite{Rav} for an overview). In this way (which will be
ours too), the NLIE stems directly from the Bethe equations and
characterises a quantum state by means of a single (or very few)
integral equation(s) in the complex plane (and possibly some
auxiliary algebraic equations). The NLIE has been widely studied
for integrable models described by trigonometric-type Bethe
equations: for instance, the 1/2-XXZ spin chain \cite{KBP}, the
inhomogeneous 1/2-XXZ and sine-Gordon field theory (ground state
in \cite{DDV}, excited states in \cite{FMQR}) and the quantum
(m)KdV/sine-Gordon theory \cite{FR1}.

In this paper, we want to propose the Non-Linear Integral Equation
idea \cite{KBP, DDV, FMQR, Rav, klump} as a tool to compute finite
$L$ corrections to the anomalous dimensions of (long) operators in
${\cal N}=4$ SYM. In terms of the solution of the NLIE, we can
indeed write down exact expressions for the {\it observable}
eigenvalues, as they depend on the Bethe roots. Their behaviour
for large $L$ may be disentangled analytically and numerically.
Going into more details, we will concentrate our analysis on the
composite operator with the highest anomalous dimension. This
corresponds in the spin chain to the {\it anti-ferromagnetic}
state, made up of a sea of real Bethe roots. We will also study
{\it excitations} \footnote{This terminology is borrowed from the
cases when the anti-ferromagnetic configuration yields the (true)
vacuum.} thereof, introduced by the presence of holes. These are
the simplest possible modifications, as already argued in
\cite{FMQR}, though the anti-ferromagnetic state is not the (true)
vacuum (with smallest energy or anomalous dimension) of the chain,
which enjoys a ferromagnetic nature and corresponds, in the gauge
theory parlance, to the BPS state with all the partons (i.e. the
complex scalars) of the same kind. On the contrary, it becomes of
interest here as its eigenvalue constitutes the upper bound, i.e.
the largest anomalous dimension: the finiteness of the spectrum is
very clear in the spin chain and SYM interpretation, although a
momentum bound of the string is rather not obvious (but
semiclassical computation can be trusted in this regime just
partially and have been started recently \cite{PTT}). Moreover, it
plays the r\^ole of the genuine vacuum in the large $N$ QCD
expansion, at least at one loop (cf., for instance, \cite{FHZ}).
Its interest resides also in the fact that the holes excitations
will furnish the just smaller anomalous dimensions, whose energies
are neglected in condensed matter physics since this part of the
spectrum decouples to infinity from the {\it real} spectrum above
the ferromagnetic vacuum, in the thermodynamic limit. In this
respect, we also emphasise that the N\'{e}el state $(|\! \!
\uparrow \downarrow \uparrow \downarrow \ldots \uparrow \downarrow
\rangle)$ is not an eigenstate in this context. Specifically, we
will study both the general $SO(6)$ case (at one loop) and the
$SU(2)$ subcase (though this with an arbitrary number of loops),
providing exact expressions for the anomalous dimensions of SYM
operators with finite number ($L$) of fields. In fact, as a
starting point for the latter we shall use the asymptotic Bethe
Ansatz of \cite{BDS}, whose reliability is up to order $g^{2L-2}$,
as already widely stressed. For clarity and simplicity reasons, we
will start by the exposition of the $SU(2)$ case which, after the
proposal by \cite{RSS}, may be interpreted as a strong coupling
expansion of the Hubbard model at {\it all} orders (provided $L$
is large enough). We will introduce the techniques in the known
example of the Heisenberg chain and we will provide original
results for the many loop Bethe Ansatz of \cite{BDS}, as well as
for the $SO(6)$ case.

\section {The $SU(2)$ case: one loop or the Heisenberg chain. \label{heis}}
\setcounter{equation}{0}

Let us first consider the $SU(2)$ subsector of the gauge-invariant
scalar operators in ${\cal N}=4$ SYM field theory. The anomalous
dimension of a general composite operator containing $L$ scalars
is given by
\begin{equation}
\gamma = \frac {\lambda }{8 \pi ^2} E \, , \label{andim}
\end{equation}
where $\lambda=Ng_{YM}^2=8\pi ^2 g^2$ is the 't Hooft coupling of
the $SU(N)$ super Yang-Mills theory and
\begin{equation}
E=\sum _{k=1}^M \frac {1}{u_k^2+\frac {1}{4}} \label{energy}
\end{equation}
is the energy of a spin $1/2$-XXX chain, i.e. the celebrated
Heisenberg spin chain, with $L$ sites. Since the pioneering work
of Bethe \cite{Bethe} it is well known that the $M$ complex
numbers (or Bethe roots) $u_k$ must satisfy the equations
\begin{equation}
\left (\frac {u_j-\frac {i}{2}}{u_j+\frac {i}{2}} \right )^L=
\mathop{\prod^M _ {k=1}}_{k\not=j} \frac {u_j-u_k-i}{u_j-u_k+i}
\label {eqbethe2} \, , \quad j=1, \ldots , M \, ,
\end{equation}
usually named after Bethe as well. In this approach, one set
$\{u_k\}$ of solutions identifies one energy eigenfunction. In the
original paper the previous equations are the consequence of the
imposition of periodicity of the postulated wave function (the
famous Ansatz), without any clear mention to integrability. This
Bethe eigenfunction is also (highest weight) eigenstate of the
total $z$-component spin operator with integer or half-integer
eigenvalue $S=L/2-M\geqslant 0$.

Now, we derive a single Non-Linear Integral Equation (NLIE) along
the ideas of \cite{DDV, FMQR}, so that we may have in it a more
effective, though equivalent, description of Bethe equations. We
may need to say that this derivation will have a pedagogical
purpose in perspective of the multi-loop case of next section,
though it will help to illustrate the general idea of \cite{DDV,
FMQR} and to interpret the results from the gauge theory
viewpoint. In fact, on the one hand it is just a limiting case of
the {\it general} Bethe Ansatz of the next section, on the other
hand similar results are already contained in \cite{KBP,klump}.

The NLIE will be an equation for the so-called counting function,
\begin{equation}
Z(x)=L\phi \left (x, \frac {1}{2} \right )- \sum _{k=1}^M \phi
(x-u_k,1) \, , \label {1Z2}
\end{equation}
where the function
\begin{equation}
\phi (x, \xi)=i \ln \left (\frac {i\xi +x }{i\xi - x} \right ) = 2
\arctan \frac {x}{\xi} \, , \quad \xi >0 \, , \label {phi}
\end{equation}
is analytic in the strip $|{\mbox {Im}}x| < \xi $ provided the
branch of the logarithm is along the negative real axis. Then we
need a variable that keeps into account the parity of the chain in
relation with the number of Bethe roots:
\begin{equation}
\label{delta} \delta= (L-M) \mbox{ mod } 2 \,.
\end{equation}
After, by using the simple property
\begin{equation}
i \ln \frac {x-i\xi}{x+i\xi}-i \ln \frac {i\xi -x}{i\xi +x}=\pi \,
, \label{logprop}
\end{equation}
the Bethe equations can be written in the form
\begin{equation}
iL \ln \frac {\frac {i}{2}+u_j} {\frac {i}{2}-u_j}-\sum _{{k=1}}^M
i \ln \frac {i+u_j-u_k}{i-u_j+u_k}=\pi (2 I_j+\delta -1) \, ,
 \quad j=1, \ldots , M \, , \nonumber
\end{equation}
thanks to the introduction of certain integer quantum numbers
$I_j$, or in terms of the counting function as
\begin{equation}\label{betheZ}
Z(u_j)=\pi (2I_j+\delta -1) \, ,  \quad j=1, \ldots , M \, .
\end{equation}
The last equations are completely equivalent to the initial Bethe
ones (\ref{eqbethe2}), provided that $u_j$ enter the counting
function by (\ref {1Z2}).

From now on and only for simplicity reasons we will be considering
states characterised by real roots. This is the formulation
proposed in \cite{FMQR} and it can be easily extended to deal with
arbitrary complex roots. Bearing in mind the limits
\begin{equation}\label{limits}
\lim _{x\rightarrow \pm \infty} \phi (x, \xi) = \pm \pi \, ,
\end{equation}
we easily compute the limiting values of the counting function
\begin{equation}
\lim _{x\rightarrow \pm \infty} Z(x) = \pm (L-M)\pi \, .\nonumber
\end{equation}
Since $Z(x)$ is an increasing function, the condition (\ref
{betheZ}) is satisfied by $L-M$ points on the real axis, among
which there are indeed $M$ Bethe roots. The number of the
remaining {\it fake} \footnote{This in the obvious sense that
these are {\bf not} solutions of the initial Bethe equations
(\ref{eqbethe2}).} solutions (holes) is
\begin{equation} \label{holes}
H=L-M-M =L-2M\,.
\end{equation}
Of course, the holes $x_h$ are determined by the same equations as
those for real roots, but with the complementary set of integer
quantum numbers $I_h$, namely
\begin{equation}\label{qholes}
Z(x_h)=\pi (2I_h+\delta -1) \, ,
\end{equation}
since holes do not satisfy the Bethe equations. Hence, both Bethe
roots and holes, respectively, enjoy the condition
\begin{equation}
{\mbox {exp}}[iZ(x)]=(-1)^{\delta-1} \, , \quad x=u_j, x_h \, .
\label {condi2}
\end{equation}
Now, let $f(u)$ be an analytic function within a strip around the
real axis. Thanks to (\ref {condi2}), the sum of its values at all
the Bethe roots takes on the expression \cite{FMQR}
\begin{eqnarray}
\sum _{k=1}^{M}f(u _k)&=&-\int_{-\infty}^{\infty }\frac{dx}{2\pi
i}f'(x-i\epsilon)\,
\ln\left[1+(-1)^{\delta}e^{iZ(x-i\epsilon)}\right] - \label {equ2} \\
&-&\int _{\infty}^{-\infty }\frac{dx}{2\pi i} f'(x+i\epsilon)\,
\ln \left[1+(-1)^{\delta}e^{iZ(x+i\epsilon)}\right] - \sum
_{h=1}^{H}f(x _h)\, , \nonumber
\end{eqnarray}
with $\epsilon > 0$ small enough to keep the integration inside
the analyticity strip. If now $\epsilon \rightarrow 0$, we may
rearrange this expression as
\begin{eqnarray}
\sum _{k=1}^{M}f(u _k)&=&-\int
_{-\infty}^{\infty}\frac{dx}{2\pi}f'(x)Z(x)+
\label{prop2}\\
&+&\int _{-\infty}^{\infty }\frac{dx}{\pi} f'(x){\mbox{ Im}}\ln
\left [1+(-1)^{\delta}e^{iZ(x+i0)}\right] - \sum _{h=1}^{H}f(x
_h)\, . \nonumber
\end{eqnarray}
Upon applying (\ref{prop2}) to the sum over the Bethe roots in the
definition of the counting function (\ref{1Z2}), we obtain yet a
first integral equation for it,
\begin{eqnarray}
Z(x)&=&L\phi \left (x, \frac {1}{2} \right )- \int _{-\infty}^{\infty}\frac {dy}{2\pi} \phi '(x-y,1)Z(y)+\nonumber \\
&+&\int _{-\infty}^{\infty }\frac {dy}{\pi} \phi '(x-y,1){\mbox {
Im}}
\ln \left [1+(-1)^{\delta}e^{iZ(y+i0)}\right] + \label{22eq1} \\
&+ & \sum _{h=1}^H \phi (x-x_h,1) \, . \nonumber
\end{eqnarray}
As usual, we introduce a shorthand
\begin{equation}
\label{elle} L(x)={\mbox {Im}} \ln \left
[1+(-1)^{\delta}e^{iZ(x+i0)}\right] \, ,
\end{equation}
and then take the Fourier transform {\footnote {We define the
Fourier transform $\hat f(k)$ of a function $f(x)$ as given by
\begin{equation}
\hat f(k)= \int _{-\infty}^{\infty} dx\ e^{-ikx} f(x) \, .
\end{equation}}}
of all terms in (\ref{22eq1}) to obtain
\begin{equation}
\hat Z(k)=L \hat \phi \left (k, \frac {1}{2} \right ) - \frac
{1}{2\pi} \hat {\phi '} (k,1)  \hat Z(k) + \frac {1}{\pi} \hat
{\phi '} (k,1)  \hat L(k) + \sum _{h=1}^H e^{-ikx_h} \hat \phi
(k,1) \, . \label {22eq2}
\end{equation}
This equation can be recast in the more compact form
\begin{equation}
\hat Z(k)=\hat F(k)+ 2 \hat G(k) \hat L(k) + \sum _{h=1}^H
e^{-ikx_h} \hat H(k)\, , \nonumber
\end{equation}
where the Fourier transform of the forcing term reads as
\begin{equation}
\hat F(k)=L \frac { \hat \phi \left (k, \frac {1}{2} \right
)}{1+\frac {1}{2\pi} \hat {\phi '} (k,1) }  \, , \label {2Fk}
\end{equation}
that of the kernel as
\begin{equation}
\hat G(k)= \frac {\frac {1}{2\pi} \hat {\phi '} (k,1) }{1+\frac
{1}{2\pi} \hat {\phi '} (k,1) } \, , \label {2Gk2}
\end{equation}
and eventually the holes contribution is ($P$ is the principal
value distribution)
\begin{equation}
\hat H(k)= \frac {\hat {\phi } (k,1) }{1+\frac {1}{2\pi} \hat
{\phi '} (k,1) }=\frac {2\pi}{i} P\left (\frac {1}{k}\right ) \hat
G(k) \, . \label {2Sk}
\end{equation}
All these can be easily calculated, once the Fourier transform of
the function
\begin{equation}
\phi '(x,\xi)=\frac {2\xi}{\xi ^2 +x^2} \, ,
\end{equation}
is explicitly computed as
\begin{equation}
\hat {\phi '} (k,\xi)=2\pi  e^ {- \xi |k|} \, , \label{phitra}
\end{equation}
which entails
\begin{equation}\label{FGk}
\hat F(k)=L  P\left (\frac {1}{k}\right ) \frac { 2\pi e^{-\frac
{|k|}{2}}}{i(1+e^{-|k|}) } \, , \quad \hat G(k)=\frac
{1}{1+e^{|k|} } \, .
\end{equation}
Upon anti-transforming, we obtain the forcing term
\begin{equation}
F(x)=L \int _{0}^{\infty} \frac {dk}{k}\, \frac {\sin kx}{\cosh
\frac {k}{2}} =2L \arctan e^{\pi x}-\frac {L\pi}{2} =L \, {\mbox
{gd}}\, \pi x \, ,\nonumber
\end{equation}
and besides the kernel
\begin{equation}
G(x)=\int_{-\infty}^{\infty} \frac {dk}{2\pi}\, e^{ikx} \frac
{1}{1+e^{|k|}} =\frac {1}{2\pi i}\frac {d}{dx}\ln \frac {\Gamma
\left ( 1+\frac {ix}{2} \right ) \Gamma \left ( \frac {1}{2}-\frac
{ix}{2} \right ) }{\Gamma \left ( 1-\frac {ix}{2} \right ) \Gamma
\left ( \frac {1}{2}+\frac {ix}{2} \right ) }= \frac {1}{2\pi
i}\frac {d}{dx}\ln S(x)\, , \nonumber
\end{equation}
where the expression in terms of Euler's gamma functions,
\begin{equation}
\quad S(x)=\frac {\Gamma \left ( 1+\frac {ix}{2} \right ) \Gamma
\left ( \frac {1}{2}-\frac {ix}{2} \right ) }{\Gamma \left (
1-\frac {ix}{2} \right ) \Gamma \left ( \frac {1}{2}+\frac {ix}{2}
\right ) }  \, ,\nonumber
\end{equation}
is indeed {\it the scattering factor} of the NLIE (cf. \cite{FMQR,
FRT, FR2} for a justification of this name). Finally, from this we
can easily gain the hole function in the form
\begin{equation}
H(x)= 2\pi \int _{0}^x dy \, G(y) \nonumber =-i\ln S(x) \,.
\end{equation}
With all these functions at hand, we are in the position to write
down the announced non-linear integral equation for $Z(x)$,
\begin{equation}
Z(x)=F(x)-i \sum _{h=1}^H \ln S(x-x_h)+2 \int
_{-\infty}^{\infty}dy\, G(x-y) {\mbox { Im}}\ln \left
[1+(-1)^{\delta}e^{iZ(y+i0)}\right]  \, ,  \label {nlin22}
\end{equation}
and we can also check {\it a posteriori} that no zero-modes
actually entered its derivation.

The NLIE (\ref{nlin22}) together with the holes quantization
conditions (\ref {qholes}) is equivalent to the Bethe equations
(\ref {eqbethe2}).

\subsection{The {\it observable} eigenvalues.}

Let us now pass on to the computation of the eigenvalues of the
{\it observables} on states containing $M$ real Bethe roots and
$H$ holes. We move from
\begin{eqnarray}
\sum _{k=1}^{M}f(u _k)&=&-\int _{-\infty}^{\infty}\frac {dx}{2\pi}
f'(x)Z(x)+
\label {prop3}\\
&+&\int _{-\infty}^{\infty }\frac {dx}{\pi} f'(x){\mbox { Im}} \ln
\left [1+(-1)^{\delta}e^{iZ(x+i0)}\right] -\sum _{h=1}^H f(x_h) \,
 \nonumber
\end{eqnarray}
and then insert into this expression the non-linear integral
equation (\ref{nlin22}) and re-organise the terms as
\begin{gather}
\sum _{k=1}^{M}f(u _k)=-\int _{-\infty}^{\infty}\frac {dx}{2\pi}
f'(x) F(x) + \sum _{h=1}^H \left\{ \int _{-\infty}^{\infty}\frac
{dx}{2\pi} f'(x)\ i \ln S(x-x_h)
- f(x_h) \right\} + \notag \\
+\int _{-\infty}^{\infty }\frac {dx}{\pi} f'(x)\int
_{-\infty}^{\infty} dy\ [\delta (x-y)-G(x-y)]{\mbox { Im}}\ln
\left [1+(-1)^{\delta}e^{iZ(y+i0)}\right] \, . \label {prop4}
\end{gather}
This formula gives an exact expression for the eigenvalues of any
general observable in terms of the solution of the Non-Linear
Integral Equation, solution which characterises the specific
eigenstate. For example, its analogue was exploited in the quantum
(m)KdV context \cite{FR1} to obtain the quintessence of an
integrable model, namely the (commuting) integrals of motion. Now,
we want to use it in order to compute the eigenvalues of the
energy (anomalous dimension) and of the momentum.

\subsection{The anomalous dimension}

As illustrated in (\ref{energy}), to compute the (total) energy,
we need to apply formula (\ref {prop4}) with the single particle
energy
\begin{equation}\label{energyfunc}
f(x)\equiv e(x)\equiv\frac {1}{x^2+\frac {1}{4}} \, .
\end{equation}
Indeed, the first term of the l.h.s. is given by
\begin{eqnarray}
-\int _{-\infty}^{\infty}\frac {dx}{2\pi} e'(x)F(x)&=&\int _{-\infty}^{\infty}\frac {dx}{2\pi} e(x)F'(x)= L \int _{-\infty}^{\infty}\frac {dx}{2} \frac  {1}{x^2+\frac {1}{4}} \frac {1}{\cosh \pi x} = \nonumber \\
&=& L \int _{-\infty}^{\infty} dy \frac {1}{y^2+1} \frac {1}{\cosh
\frac {\pi y}{2}}= 2 L \ln 2 \, .  \label {e-force}
\end{eqnarray}
The last term reads
\begin{equation}
\int _{-\infty}^{\infty }\frac {dx}{\pi} e'(x)\int
_{-\infty}^{\infty} dy\ [\delta (x-y)-G(x-y)]{\mbox { Im}}\ln
\left [1+(-1)^{\delta}e^{iZ (y+i0)}\right]  \, ,\nonumber
\end{equation}
where the $x$-convolution is conveniently evaluated in Fourier
space (where it becomes an ordinary product):
\begin{gather}
 \int _{-\infty}^{\infty }\frac {dy}{\pi} {\mbox { Im}}\ln \left [1+(-1)^{\delta}e^{iZ(y+i0)}\right] \frac {d}{dy} \int _{-\infty}^{\infty }dk \frac {e^{iky}}{2\cosh \frac {k}{2}}=\notag  \\
=  \int _{-\infty}^{\infty }dy \left( \frac {d}{dy}\ \frac
{1}{\cosh {\pi y}}\right) {\mbox {Im}}\ln \left
[1+(-1)^{\delta}e^{iZ(y+i0)}\right] \label {e-bulk} \, .
\end{gather}
Eventually, we need to compute the two terms of the hole sum (the
second term):
\begin{equation}
\sum _{h=1}^H \left\{ \int _{-\infty}^{\infty}\frac {dx}{2\pi}
e'(x)\ i \ln S(x-x_h) - e(x_h)\right\} \, .\nonumber
\end{equation}
For the first of them we may write
\begin{equation}
\int _{-\infty}^{\infty}\frac {dx}{2\pi}\, e'(x)\, i \ln S(x-x_h)
= \int _{-\infty}^{\infty} dx\, e(x)\, G(x-x_h) = \int
_{-\infty}^{\infty}\frac {dk}{2\pi}\, e^{ikx_h}\, \hat e(k)\, \hat
G(k)\, .\nonumber
\end{equation}
This yields, once the second term is expressed by its Fourier
transform,
\begin{equation}
\sum _{h=1}^H \int _{-\infty}^{\infty}\frac {dk}{2\pi}\,
e^{ikx_h}\, \hat e(k)\, [\hat G(k) -1] = - \sum _{h=1}^H \int
_{-\infty}^{\infty} dk\, \frac {e^{ikx_h}}{2 \cosh \frac {k}{2}}
\, , \label{e-source1}
\end{equation}
where we have used the formula
\begin{equation}
\hat e(k)= 2\pi e^{-\frac {1}{2}|k|} \, ,\nonumber
\end{equation}
particular case of (\ref{phitra}), and the expression of $\hat
G(k)$, (\ref{FGk}). Eventually, the source term may be written as
\begin{equation}
 - \sum _{h=1}^H \int _{-\infty}^{\infty} dk\, \frac {e^{ikx_h}}{2 \cosh \frac {k}{2}} =
- \sum _{h=1}^H \frac {\pi}{\cosh \pi x_h} \, . \label{e-source2}
\end{equation}
Summing up all the contributions (\ref {e-force}, \ref {e-bulk},
\ref {e-source2}), for the eigenvalue of the energy of the spin
chain, we obtain
\begin{eqnarray}
E &=& 2L\ln 2 - \sum _{h=1}^H  \frac {\pi}{\cosh \pi x_h} + \nonumber \\
&+& \int _{-\infty}^{\infty }dy \left(\frac {d}{dy} \frac
{1}{\cosh {\pi y}}\right) {\mbox {Im}}\ln \left
[1+(-1)^{\delta}e^{iZ(y+i0)}\right]  \, . \label{gammasu2}
\end{eqnarray}
This expression is exact for any $L$ and gives the largest
anomalous dimensions of the gauge-invariant scalar operators of
the $SU(2)$ sub-sector in ${\cal N}=4$ SYM. The first term on the
r.h.s. is the known leading term proportional to $L$; the
remaining two addends may be expanded in the limit $L\rightarrow
\infty$ to provide respectively ${\cal O}(1)$ and ${\cal O}(1/L)$
corrections. Analytical expressions up to the order $1/L$ will be
given in Section 5.

\subsection{The momentum}
The identification between the anomalous dimension of a
gauge-invariant operator and the energy of a (spin chain) state
needs to be supplemented by the zero momentum condition.
Therefore, it is necessary to work out, by using the same
technology as for the energy, the momentum eigenvalue \footnote{Of
course, the identification up to $2\pi$ multiples comes from its
definition as a displacement operator on a periodic
(one-dimensional) lattice.}
\begin{equation}
\label{momentum} P=\left(\sum_{k=1}^{M} p(u_k)\right)  \
\text{mod} \  2\pi \, ,
\end{equation}
with the single particle momentum, defined as \footnote{We need to
extend the definition of the $\text{sign}$-function so that
$\text{sign}(0)=1$.}
\begin{equation}
p(x)=\frac {1}{i}\, \ln \frac {x+\frac {i}{2}}{x-\frac
{i}{2}}=\pi\, \text{sign}(x) -2 \arctan 2x =\pi\, \text{sign}(x) -
\phi \left (x, \frac {1}{2} \right ) \, . \label {px}
\end{equation}
This relation and the analogous (\ref{energyfunc}) suggest the
interpretation of each root as a particle (magnon) exciting the
ferromagnetic vacuum and obeying the energy-momentum dispersion
relation
\begin{equation}
e(u_k)=4\sin^2\frac{p(u_k)}{2}\,, \qquad k=1,...,M. \nonumber
\end{equation}
We remark that $p(x)$ is odd and discontinuous in zero
\begin{equation}
p(x)+p(-x)=0 \, , \qquad \lim _{x\rightarrow 0^{\pm}} p(x)=\pm \pi
\, .\nonumber
\end{equation}
The total momentum may be arranged so as to extract its
non-analytic contribution
\begin{equation}\label{heismom}
P=\pi (M_R^{+}-M_R^{-})-\sum_{k=1}^{M}
\phi\left(u_k,\frac12\right)= \pi M-\sum_{k=1}^{M}
\phi\left(u_k,\frac12\right) \, ,
\end{equation}
with $M_R^{+}$ the number of positive or zero real roots and
$M_R^{-}$ that of negative roots. The second equality is obtained,
modulo $2\pi$, by adding $2\pi M_R^{-}$. Now, we can apply formula
(\ref {prop4}) to the analytic part of $p(x)$,
$p_{\text{an}}(x)=-\phi\left(x,\frac12\right)$. The first term
vanishes in that the integrand is an odd function
\begin{equation}
-\int _{-\infty}^{\infty}\frac {dx}{2\pi} \,
p_{\text{an}}'(x)F(x)= 0 \, . \label{p-force}
\end{equation}
For what concerns the contribution
\begin{equation}
\int _{-\infty}^{\infty }\frac {dx}{\pi} \, p_{\text{an}}'(x)\int
_{-\infty}^{\infty} dy \, [\delta (x-y)-G(x-y)]{\mbox { Im}}\ln
\left [1+(-1)^{\delta}e^{iZ(y+i0)}\right]\, ,\nonumber
\end{equation}
we first evaluate the integration on $x$,
\begin{equation}\label{cosh}\begin{array}{c}
\displaystyle\int _{-\infty}^{\infty }dx\, p_{\text{an}}'(x)
[\delta (x-y)-G(x-y)]
=-\int _{-\infty}^{\infty }dk\, e^{iky}  \frac {e^{\frac {|k|}{2}}}{1+e^{|k|}}=\\
\displaystyle=-\int _{-\infty}^{\infty }dk\ \frac {e^{iky}}{2\cosh
\frac {k}{2}} =-\frac{\pi}{\cosh{\pi y}} \, ,
\end{array}
\end{equation}
in order to obtain
\begin{eqnarray}
&\displaystyle \int _{-\infty}^{\infty }\frac {dx}{\pi}\,
p_{\text{an}}'(x) \int _{-\infty}^{\infty} dy\, [\delta
(x-y)-G(x-y)]{\mbox { Im}}\ln \left
[1+(-1)^{\delta}e^{iZ(y+i0)}\right]
=\nonumber& \\
&\displaystyle= - \int _{-\infty}^{\infty }dy \, \frac {1}{\cosh
{\pi y}} {\mbox { Im}} \ln \left
[1+(-1)^{\delta}e^{iZ(y+i0)}\right]  \, . \label{p-bulk}&
\end{eqnarray}
Finally, we need to compute the term
\begin{equation}
\sum _{h=1}^H \left\{ \int _{-\infty}^{\infty}\frac {dx}{2\pi} \,
p_{\text{an}}'(x)\, i \ln S(x-x_h) - p_{\text{an}}(x_h)\right\} \,
. \label{source}
\end{equation}
But this expression is the sum of single hole contributions which
are minus the primitive of (\ref{cosh}) at the value $y=x_h$. So,
each term is given by
\begin{equation}
\int \frac{\pi}{\cosh\pi y} =\arctan \sinh \pi y
+\text{const}.\nonumber
\end{equation}
And the integration constant is zero since the single term has to
vanish for $y=x_h=0$ for parity reasons ($p_{\text{an}}$ is odd
and $G$ even). So (\ref{source}) simplifies to
\begin{equation}\label{sourceh}
\sum _{h=1}^H \left\{ \int _{-\infty}^{\infty}\frac {dx}{2\pi} \,
p_{\text{an}}'(x)\ i \ln S(x-x_h) - p_{\text{an}}(x_h)\right\} =
\sum _{h=1}^H \arctan\, \sinh \pi x_h \, .
\end{equation}
All the contributions (\ref {p-force}, \ref {p-bulk}, \ref
{sourceh}) yield the momentum eigenvalue
\begin{equation}
P=\pi M + \sum _{h=1}^H \left (\arctan \sinh \pi x_h \right )
-\int _{-\infty}^{\infty }dy \frac {1}{\cosh {\pi y}} {\mbox{ Im}}
\ln \left [1+(-1)^{\delta} e^{iZ(y+i0)}\right]   \, . \label
{p-eig}
\end{equation}
The attentive reader will find similar results here and there in
\cite{KBP, klump}.

As we said at the beginning of this subsection, the condition
$P=0$ works as a constraint for the anomalous dimension
(\ref{gammasu2}, \ref{andim}). In particular, the
antiferromagnetic state simply enjoys
\begin{equation}
P=\left(\pi \frac {L}{2}\right) \text{ mod }2\pi = \left \{
\begin{array}{c@{~ \text{if}~ }l} 0 &
  L\in 4{\mathbb {N}} \, , \\
\pi & L\in 4  {\mathbb {N}} +2 \, , \end{array} \right.
\label{pantif}
\end{equation}
so it possesses a SYM operator as a counterpart only if $ L\in
4{\mathbb {N}}$.

\section{The $SU(2)$ case: multi-loops}
\setcounter{equation}{0}

Following the line of the previous Section, we want to establish
the NLIE framework for the conjectured multi-loop Bethe equations
\cite{BDS}
\begin{equation}
\label{BAloops}
\left(\frac{X(u_j+\frac{i}2)}{X(u_j-\frac{i}2)}\right)^L=
\mathop{\sum^M_{k=1}}_{k\neq j} \frac{u_j-u_k+i}{u_j-u_k-i} \, ,
\end{equation}
where we introduced the function \be\label{ics} X(x)=\frac{x}2
\left( 1+\sqrt{1-\frac{\lambda}{4\pi^2 x^2}} \right). \ee And as
usual the single particle momentum $p(u_j,\lambda )$ is such that
$e^{i\, p(u_j,\lambda )\, L}$ equals the l.h.s. of the
corresponding Bethe equation (\ref{BAloops}), or explicitly
\be\label{momloops} p(x,\lambda )=\frac1{i}\ln
\frac{X(x+\frac{i}2)}{X(x-\frac{i}2)}\,. \ee Of course, these
equations reproduce the Heisenberg case of Section~\ref{heis} in
the small coupling limit $\lambda\rightarrow 0$\,: actually, only
the l.h.s. in (\ref{BAloops}), namely the momentum
(\ref{momloops}), has changed from the XXX case. Therefore, in the
present section we shall systematically follow all computations of
the previous one.

In other words, we simply need to change the function
$\phi(x,\frac12)$ of (\ref{1Z2}) into \be\label{Phi}
\Phi(x,\lambda)=i\, \ln
\frac{(\frac{i}2+x)\sqrt{1-\frac{\lambda}{4\pi^2
(x+\frac{i}{2})^2}}} {(\frac{i}2-x)\sqrt{1-\frac{\lambda}{4\pi^2
(x-\frac{i}{2})^2}}} \, , \ee which, contrarily to the momentum
(\ref{momloops}), is continuous in $x= 0$. Then, the counting
function may be defined as \be\label{Zloops} Z(x,\lambda)=L\,
\Phi(x,\lambda)-\sum _{k=1}^M \phi (x-u_k,1)\, , \ee so that the
Bethe equations read (with certain integer quantum numbers $I_j$)
\begin{gather}
Z(u_j,\lambda)=\pi (2I_j+\delta-1) \, , \label{quloops}\\
\delta\equiv(L-M)\text{ mod } 2  \, . \notag
\end{gather}
The asymptotic limits of the counting function are again given by
\be \lim_{x\rightarrow \pm\infty}Z(x,\lambda)=\pm(L-M)\, \pi
\nonumber \ee and can be used to fix the number of holes. Indeed,
as $Z(x,\lambda)$ is an increasing function, the conditions
(\ref{quloops}) with generic integers $I_j$ are at most satisfied
by $L-M$ points on the real axis. This means that the number of
holes is \be \label{holesloops} H=L-2M \, , \ee when considering
states with real roots only. The position of any hole $x_h$ is
fixed by a quantisation condition identical to (\ref{quloops}),
but with a {\it fake} quantum number $I_h$  \be Z(x_h,\lambda)=\pi
(2I_h+\delta-1) \label{qu_holes}\,. \ee Consequently, both the
Bethe roots and the holes, viz. $x=u_j, x_h$, satisfy the
condition \be\label{condloops} \exp[iZ(x,\lambda)]=(-1)^{\delta-1}
\, . \ee Again, as $\epsilon \rightarrow 0$ the sum over all the
real roots (\ref{equ2}) takes on the form
\begin{eqnarray}
\sum _{k=1}^{M}f(u _k)&=&-\int _{-\infty}^{\infty}\frac{dy}{2\pi}
\, f'(y)\, Z(y,\lambda)+
\label {prloops}\\
&+&\int _{-\infty}^{\infty }\frac{dy}{\pi}\, f'(y){\mbox { Im}}
\ln \left [1+(-1)^{\delta}e^{iZ(y+i0,\lambda)}\right] - \sum
_{h=1}^{H}f(x _h)\, . \nonumber
\end{eqnarray}
This expression may be applied to the sum in the counting function
(\ref{Zloops}) and brings
\begin{eqnarray}
Z(x,\lambda)&=&L\, \Phi(x,\lambda)- \int _{-\infty}^{\infty}\frac
{dy}{2\pi} \, \phi'(x-y,1) \,
Z(y,\lambda)+\nonumber \\
&+&\int _{-\infty}^{\infty }\frac {dy}{\pi}\, \phi '(x-y,1){\mbox
{ Im}}
\ln \left [1+(-1)^{\delta}e^{iZ(y+i0,\lambda)}\right] + \label{prog} \\
&+ & \sum _{h=1}^H \phi (x-x_h,1) \, .\nonumber
\end{eqnarray}
It is convenient to introduce the usual (cf. (\ref{elle}))
synthetic notation \be L(x,\lambda)={\mbox { Im}}\ln \left
[1+(-1)^{\delta}e^{iZ(x+i0,\lambda)}\right]\, . \nonumber \ee
After $x$ Fourier transforming all the terms and moving the first
convolution to the l.h.s., we obtain
\begin{equation}\label{Zfourier}
\hat Z(k,\lambda)=\hat F(k,\lambda)+ 2 \hat G(k) \hat L(k,\lambda)
+ \sum _{h=1}^H e^{-ikx_h} \hat H(k)\, .
\end{equation}
All terms are the same as before in the Heisenberg chain, except
the forcing term that now depends on $\lambda$ and whose Fourier
transform reads \be \hat F(k,\lambda)=L \frac { \hat \Phi (k,
\lambda )}{1+\frac {1}{2\pi} \hat {\phi '} (k,1) }  \, . \label
{forcing} \ee
\newcommand{\bessel}[1]{J_{#1}\!\left(\frac{\sqrt{\lambda}}{2\pi}k\right)}
The $x$ Fourier transform of $\Phi'(x,\lambda)$ is given in terms
of the Bessel function of the first kind $J_0$ \cite{GRA} \be
\hat\Phi'(k,\lambda)=2\pi\ e^{-\frac{|k|}2}\bessel{0} \,.\nonumber
\ee The series expansion
$\bessel{0}=(1-\frac{k^2}{16\pi^2}\lambda+{\cal O}(\lambda ^2))$
\cite{GRA} shows clearly the change with respect to the Heisenberg
chain. The function $\phi(x,1)$ is unchanged, so we can make use
of (\ref{phitra}) to arrive at the final expression
\be\label{forza} F(x,\lambda)=L \int _{-\infty}^{\infty}{dk}\,
\frac{\sin kx \, \bessel{0} }{k\ 2\cosh \frac{k}2 }=L \left (
{\mbox {gd }}\pi x - \frac {\lambda}{16}\frac {\sinh \pi x}{\cosh
^2 \pi x }+{\cal O}(\lambda ^2)\right ) \,. \ee Inverting the
Fourier transforms of (\ref{Zfourier}) leads to the NLIE valid for
this multi-loop Bethe equations \footnote{We need to stress anew
the absence of an integrable model behind them.} \ba
Z(x,\lambda)&=&F(x,\lambda)-i \sum _{h=1}^H \ln S(x-x_h)+ \label{nlieloops}\\
&+&2 \int _{-\infty}^{\infty} dy\ G(x-y)\ {\mbox {Im}}\ln \left
[1+(-1)^{\delta}e^{iZ(y+i0,\lambda)}\right]  \nonumber  \,. \ea Of
course, the convolution kernel $G$ and the hole term $S$ are the
same as in Section~\ref{heis}: what makes the difference is {\it
simply} the different forcing term $F$. And besides the structure
of this NLIE is quite the same as in many other models, except for
the specific form of the above-computed functions: hence this
similarity corroborates straight away the effectiveness of our
method.

Therefore, we can still follow the result (\ref{prop4}) on the
Heisenberg chain, keeping in mind that here the forcing term is
given by (\ref{forza}):
\begin{gather}
\sum _{k=1}^{M}f(u _k)=-\int _{-\infty}^{\infty}\frac {dx}{2\pi}
f'(x) F(x,\lambda) +  \label{fmultiloops}\\
+\int _{-\infty}^{\infty }\frac {dx}{\pi} f'(x)\int
_{-\infty}^{\infty} dy\ [\delta (x-y)-G(x-y)]{\mbox { Im}}\ln
\left [1+(-1)^{\delta}e^{iZ(y+i0, \lambda)}
\right] + \notag \\
+ \sum _{h=1}^H \left\{\int_{-\infty}^{\infty}\frac{dx}{2\pi}
f'(x)\ i \ln S(x-x_h) - f(x_h) \right\} \, .  \notag
\end{gather}

\subsection{Anomalous dimension}

As typical in Bethe Ansatz theory, the energy of the spin chain,
and thus the anomalous dimension in gauge theory, is given by a
sum on all the Bethe roots \be\label{energ} E=\sum_{j=1}^{M}
e(u_j,\lambda) \, , \ee where the (even) single particle energy
function equals \be\label{energ0}
e(x,\lambda)=i\left\{\frac1{X(x+\frac{i}2)}-\frac1{X(x-\frac{i}2)}\right\}
\,. \ee So, we just need to insert this function into
(\ref{fmultiloops})
\begin{gather}
\sum _{k=1}^{M}e(u _k,\lambda)=-\int _{-\infty}^{\infty}\frac {dx}{2\pi}
e'(x,\lambda)
F(x,\lambda) + \label{pr4loops} \\
+\int _{-\infty}^{\infty }\frac {dx}{\pi} e'(x,\lambda)\int
_{-\infty}^{\infty} dy\ [\delta (x-y)-G(x-y)]{\mbox { Im}}\ln
\left [1+(-1)^{\delta}e^{iZ(y+i0, \lambda)}
\right] + \notag \\
+ \sum _{h=1}^H \left\{\int_{-\infty}^{\infty}\frac{dx}{2\pi}
e'(x,\lambda)\ i \ln S(x-x_h) - e(x_h,\lambda) \right\} \, ,  \notag
\end{gather}
and re-call its Fourier transform \be \hat
e(k,\lambda)=\frac{8\pi^2\bessel{1}}{\sqrt{\lambda}\ k\
e^{\frac{|k|}2}}\,.\nonumber \ee In fact, the first contribution
reads
\begin{gather}
-\int _{-\infty}^{\infty}\frac{dx}{2\pi}\ e'(x,\lambda)\,
F(x,\lambda)=
\int _{-\infty}^{\infty}\frac{dx}{2\pi}\ e(x,\lambda)\, F'(x,\lambda)=\notag \\
=\frac{8\pi}{\sqrt{\lambda}}L \int _{0}^{\infty} dk\
\frac{\bessel{0}\bessel{1}}{k\, (e^k+1)}= L\left ( 2\ln
2-\frac{9\zeta(3)}{8(2\pi)^2}  \lambda + {\cal
{O}}(\lambda^2)\right) \,. \label{energ1}
\end{gather}
Of course, it is the leading term of the anti-ferromagnetic state
energy and thus coincides with the expression (10) of \cite{RSS}
(or (17,18) of \cite{ZAR}), where it was interestingly identified
with the ground state energy of the half-filled Hubbard model. Moreover,
the second contribution in (\ref{pr4loops}) may be re-organised by
expressing the convolution as an ordinary product in the Fourier
space:
\begin{gather}
\int _{-\infty}^{\infty }\frac {dx}{\pi} \, e'(x,\lambda)\int
_{-\infty}^{\infty} dy\, [\delta (x-y)-G(x-y)]{\mbox { Im}}\ln
\left [1+(-1)^{\delta}e^{iZ(y+i0, \lambda)}
\right]= \notag\\
=\int _{-\infty}^{\infty }dy\, e_1(y,\lambda) {\mbox { Im}}\ln
\left [1+(-1)^{\delta}e^{iZ(y+i0, \lambda)}\right]\, , \label{energ2}
\end{gather}
where we made use of a new function as a Fourier transform of a
construct of the Bessel function $J_1$
\begin{gather}\label{e1}
e_1(x,\lambda)= \int _{-\infty}^{\infty }dk\, \frac{2i\bessel{1}
e^{ikx}}{\sqrt{\lambda}\, \cosh \frac{k}2} =
\\
=\frac{d}{dx}\left[ \frac1{\cosh \pi x}+\frac{\lambda}{32 \pi^2}\
\frac{d^2}{dx^2}\ \frac1{\cosh \pi x} +{\cal {O}}(\lambda^2)
\right] \, . \notag
\end{gather}
The single hole contribution in (\ref{pr4loops}) is now evaluated
by integrating by parts and computing the convolution in the
Fourier space:
\begin{gather}
\int_{-\infty}^{\infty}\frac{dx}{2\pi}\, e'(x,\lambda)\ i \ln
S(x-x_h) - e(x_h,\lambda) = \int_{-\infty}^{\infty} dx\
e(x,\lambda)\,
[G(x-x_h) - \delta(x-x_h)] = \notag \\
=\int_{-\infty}^{\infty} \frac{dk}{2\pi}\ e^{ikx_h}\, \hat
e(k,\lambda)\ [\hat G(k)-1]=-\int_{-\infty}^{\infty} dk\, 2\pi\,
e^{ikx_h}\, \frac{\bessel{1}}
{\sqrt{\lambda}\ k \cosh \frac{k}2}= \notag \\
=-\frac{\pi}{\cosh \pi x_h} -\frac{\lambda}{32 \pi^2}
\left[\frac{d^2}{dx^2}\ \frac{\pi}{\cosh \pi x}\right]_{x=x_h} +
{\cal{O}}(\lambda^2) \, . \label{energ3}
\end{gather}
Eventually, we collect all terms (\ref{energ1}, \ref{energ2},
\ref{energ3}) that form the energy (\ref{energ}) and obtain
\begin{gather}
E=\frac{8\pi}{\sqrt{\lambda}}L
\int _{0}^{\infty} dk\ \frac{\bessel{0}\bessel{1}}{k(e^k+1)} - \label{energ4} \\
-\sum_{h=1}^H \int_{-\infty}^{\infty} dk\ 2\pi\, e^{ikx_h}\,
\frac{\bessel{1}}
{\sqrt{\lambda}\ k \cosh \frac{k}2} + \notag \\
+\int _{-\infty}^{\infty }dy\ \left( \int _{-\infty}^{\infty }dk\
\frac{2i\bessel{1} e^{iky}}{\sqrt{\lambda}\ \cosh \frac{k}2}
\right) {\mbox { Im}}\ln \left
[1+(-1)^{\delta}e^{iZ(y+i0,\lambda)}\right] \notag\,.
\end{gather}

This expression for the anomalous dimension is exact in any
regime of $L$ and specifically the second and third terms provide
all the sub-leading corrections when $L\rightarrow \infty$, whose
expressions are by the same token novel and intriguing. Analytical
expressions of them up to the order $1/L$ will be given in Section
5.

In conclusion, it is worth emphasising that the break-down of the
proposal (\ref{BAloops}) at order $g^{2L}$ affects our results as
a trivial consequence, although it reveals itself unrelated to our
method. In other words, the latter should be perfectly applicable
to the hypothetic correct Bethe Ansatz equations, provided in the
{\it typical} form, along similar steps.

\subsection{Momentum}
We can compute the momentum (\ref{momentum}) by summing the single
particle momenta (\ref{momloops}) of all the Bethe roots. From
(\ref{momloops}) we separate the analytic contribution
$p_{\text{an}}(u_k,\lambda)=-\Phi(u_k,\lambda)$ so that, as in
(\ref{heismom}), we can write \be P=\pi M -\sum_{k=1}^M
\Phi(u_k,\lambda)\,. \ee Thus, we only need to apply
(\ref{fmultiloops}) to the function $p_{an}(x,\lambda)$. The first
contribution vanishes \be\label{enlo1} -\int _{-\infty}^{\infty
}\frac{dx}{2\pi}\, p_{\text{an}}'(x,\lambda) \, F(x,\lambda)=0 \ee
as the integrand is the product of an even and an odd function.
The second contribution to (\ref{fmultiloops}) becomes easily
\begin{gather}
\int _{-\infty}^{\infty }\frac {dx}{\pi}\,
p_{\text{an}}'(x,\lambda) \int _{-\infty}^{\infty} dy\, [\delta
(x-y)-G(x-y)]{\mbox { Im}}\ln \left [1+(-1)^{\delta}e^{iZ(y+i0,
\lambda)}\right]
=\notag  \\
= - \int _{-\infty}^{\infty }\frac{dy}{\pi} \int
_{-\infty}^{\infty}dk\, \frac{e^{iky}\bessel{0} }{2\cosh \frac{k}2
}
{\mbox { Im}}\ln \left [1+(-1)^{\delta}e^{iZ(y+i0, \lambda)}\right] = \notag \\
= - \int _{-\infty}^{\infty }\frac{dy}{\pi}\,
\frac{F'(y,\lambda)}{L} {\mbox { Im}}\ln \left
[1+(-1)^{\delta}e^{iZ(y+i0, \lambda)}\right] \, , \label{ploop}
\end{gather}
with the appearance of the forcing/momentum term (\ref{forza}).
The latter also appears in the hole contribution to
(\ref{fmultiloops})
\begin{gather}
\int _{-\infty}^{\infty}\frac {dx}{2\pi} \,
p_{\text{an}}'(x,\lambda)\, i \ln S(x-x_h) - p(x_h,\lambda) =  \notag \\
=\int _{-\infty}^{\infty} dx\, p_{\text{an}}(x,\lambda)\,
[G(x-x_h)-\delta(x-x_h)] =\frac{F(x_h,\lambda)}{L} \, .
\label{sourceloops}
\end{gather}
In summary, we have the analogue of (\ref{p-eig}) in the present
case
\begin{equation}
P=\pi M + \frac1{L} \sum _{h=1}^H F(x_h,\lambda) -\int
_{-\infty}^{\infty }\frac{dy}{\pi}\, \frac {F'(y,\lambda)}{L}
{\mbox{ Im}} \ln \left [1+(-1)^{\delta} e^{iZ(y+i0,
\lambda)}\right] \, . \label {p-eigloops}
\end{equation}
This completes the general results of the multi-loop scenario and
will allow us to extract in Section 5 the first finite size
corrections analytically (and explicitly).

\section {The $SO(6)$ scalar sector at one loop: finite size results}
\setcounter{equation}{0}

We want to illustrate the utility of the NLIE to compute the exact
finite size contributions to the anomalous dimensions and momenta
in the $SO(6)$ scalar sector at one loop. Therefore, we need to
consider a chain of $L$ six-dimensional vectors of the $so(6)$
representation. As well known after \cite{resdev}, the Bethe
Ansatz diagonalization of all the commuting integrals of motion is
founded on the following system of coupled Bethe equations:
\begin{eqnarray}
\left ( \frac {u_{1,j}+i/2}{u_{1,j}-i/2}\right )^L&=&
\mathop{\prod _{k=1}^{M_1}}_{k\neq j} \frac
{u_{1,j}-u_{1,k}+i}{u_{1,j}-u_{1,k}-i}
\prod _{k=1}^{M_2} \frac {u_{1,j}-u_{2,k}-i/2}{u_{1,j}-u_{2,k}+i/2}\prod _{k=1}^{M_3} \frac {u_{1,j}-u_{3,k}-i/2}{u_{1,j}-u_{3,k}+i/2} \, , \nonumber \\
1&=&\mathop{\prod _{k=1}^{M_2}}_{k\not = j} \frac {u_{2,j}-u_{2,k}+i}{u_{2,j}-u_{2,k}-i}\prod _{k=1}^{M_1} \frac {u_{2,j}-u_{1,k}-i/2}{u_{2,j}-u_{1,k}+i/2} \, , \\
1&=&\mathop{\prod _{k=1}^{M_3}}_{k\not = j} \frac
{u_{3,j}-u_{3,k}+i}{u_{3,j}-u_{3,k}-i}\prod _{k=1}^{M_1} \frac
{u_{3,j}-u_{1,k}-i/2}{u_{3,j}-u_{1,k}+i/2} \, .\nonumber
\end{eqnarray}
By making use of the function (\ref{phi}), we may define three
counting functions, i.e. one for each group of Bethe equations,
\begin{eqnarray}
Z_1(u)&=&L\, \phi  (u,  1/2  ) - \sum
_{k=1}^{M_1} \phi (u-u_{1,k},1) + \nonumber \\
&+&\sum _{k=1}^{M_2} \phi (u-u_{2,k},1/2)+\sum _{k=1}^{M_3} \phi
(u-u_{3,k},1/2) \, ,
\nonumber \\
Z_2(u)&=&-\sum _{k=1}^{M_2} \phi (u-u_{2,k},1)+\sum _{k=1}^{M_1} \phi (u-u_{1,k},1/2) \, , \\
Z_3(u)&=&-\sum _{k=1}^{M_3} \phi (u-u_{3,k},1)+\sum _{k=1}^{M_1}
\phi (u-u_{1,k},1/2) \, , \nonumber
\end{eqnarray}
such that the Bethe equations look as
\begin{eqnarray}
Z_1(u_{1,j})&=&\pi (2I_{1,j}+\delta_1-1)  \, , \nonumber \\
Z_2(u_{2,j})&=&\pi (2I_{2,j}+\delta_2-1) \, , \\
Z_3(u_{3,j})&=&\pi (2I_{3,j}+\delta_3-1) \, , \nonumber
\end{eqnarray}
where $I_{k,i}$ are integer quantum numbers and the (\ref{delta})
has been generalised to \ba
\delta_1&=&(L-M_1+M_2+M_3)\text{ mod } 2 \,, \nonumber\\
\delta_2&=&(M_1-M_2)\text{ mod } 2\,,\\
\delta_3&=&(M_1-M_3)\text{ mod } 2\,. \nonumber \ea We now
consider states of the chain described by real solutions $\{
u_{k,i} \}$ to the Bethe equations. For simplicity's sake, we
consider the case in which the parities of the integers
$L,M_1,M_2,M_3$ are such that ${\mbox {exp}}[iZ_k(u_{k,i})]=-1$
(compare with (\ref{condi2})). In addition, even if the formalism
would allow us to consider states with holes (and generally
complex roots), here we limit ourselves to states which contain no
holes, i.e. such that the points $u$ satisfying ${\mbox
{exp}}[iZ_k(u)]=-1$ are exhausted by a (real) solution set $\{
u_{k,i} \}$: we will be extending our results in an incoming
publication \cite{35}. As before, these requirements will
constrain the allowed values of $M_k$. Indeed, from the limits
(\ref{limits}) we obtain the limiting values
\begin{eqnarray}
\lim _{x\rightarrow \pm \infty} Z_1(x)&=&\pm \pi (L-M_1+M_2+M_3) \, , \nonumber \\
\lim _{x\rightarrow \pm \infty} Z_2(x)&=&\pm \pi (M_1-M_2) \, , \nonumber \\
\lim _{x\rightarrow \pm \infty} Z_3(x)&=&\pm \pi (M_1-M_3) \, .
\nonumber
\end{eqnarray}
Imposing the condition that the points $u$ satisfying ${\mbox
{exp}}[iZ_k(u)]=-1$ are those and only those in the solution set
$\{ u_{k,i} \}$ furnishes these constraints
\begin{equation}
|L-M_1+M_2+M_3|=M_1 \, , \quad |M_1-M_2|=M_2 \, , \quad
|M_1-M_3|=M_3 \, ,
\end{equation}
whose solution, if $M_1\neq 0$, is
\begin{equation}
M_1=L \, , \quad M_2=M_3=\frac {L}{2} \, .
\end{equation}
If $L\in 4\mathbb{N}$ there is one single state with these
features and it is the completely anti-ferromagnetic state: as
discussed in \cite{MZ} it is the state with maximal energy and
zero momentum, too. Now, the usual procedure allows us to write a
sum over the Bethe roots of a function $f$ (analytic within a
strip around the real axis) in terms of integrals involving the
$Z$s:
\begin{equation}
\sum _{i=1}^{M{_k}}  f(u_{k,i})=-\int _{-\infty}^{\infty} \frac
{dx}{2\pi} \, f^{'}(x) Z_k(x) +\int _{-\infty}^{\infty} \frac
{dx}{\pi} \, f^{'}(x) {\mbox { Im}} \ln \left [
1+e^{iZ_k(x+i0)}\right ] \, . \label {sumf}
\end{equation}
Upon applying (\ref{sumf}) in the definition of $Z_k$, we
are on the road to write NLIEs for the counting functions:
\begin{eqnarray}
&&Z_1(x)=L\, \phi (x,1/2)-\int _{-\infty}^{\infty} \frac
{dy}{2\pi}
\,  \phi^{'}(x-y,1) \, Z_1(y) + \nonumber \\
&+&2 \int _{-\infty}^{\infty} \frac {dy}{2\pi}\, \phi ^{'}(x-y,1)\, {\mbox {Im}} \ln \left [ 1+e^{iZ_1(y+i0)}\right ] + \nonumber  \\
&+&\int _{-\infty}^{\infty} \frac {dy}{2\pi}\, \phi
^{'}(x-y,1/2)\, Z_2(y) - 2 \int _{-\infty}^{\infty} \frac
{dy}{2\pi}\, \phi ^{'}(x-y,1/2)\,
{\mbox {Im}} \ln \left [ 1+e^{iZ_2(y+i0)}\right ] + \nonumber  \\
&+&\int _{-\infty}^{\infty} \frac {dy}{2\pi} \, \phi
^{'}(x-y,1/2)\, Z_3(y) - 2 \int _{-\infty}^{\infty} \frac
{dy}{2\pi} \, \phi ^{'}(x-y,1/2)\, {\mbox {Im}} \ln \left [
1+e^{iZ_3(y+i0)}\right ] \, , \nonumber
\end{eqnarray}
\begin{eqnarray}
Z_2(x)&=& -\int _{-\infty}^{\infty} \frac {dy}{2\pi}\, \phi
^{'}(x-y,1) \, Z_2(y) + 2 \int _{-\infty}^{\infty} \frac
{dy}{2\pi}\, \phi ^{'}(x-y,1)
\, {\mbox {Im}} \ln \left [ 1+e^{iZ_2(y+i0)}\right ] + \nonumber \\
&+&\int _{-\infty}^{\infty} \frac {dy}{2\pi}\, \phi ^{'}(x-y,1/2)
\, Z_1(y) - 2 \int _{-\infty}^{\infty} \frac {dy}{2\pi} \, \phi
^{'}(x-y,1/2)\,  {\mbox {Im}} \ln \left [ 1+e^{iZ_1(y+i0)}\right ]
\, , \nonumber
\end{eqnarray}
\begin{eqnarray}
Z_3(x)&=& -\int _{-\infty}^{\infty} \frac {dy}{2\pi}\, \phi
^{'}(x-y,1) \, Z_3(y) + 2 \int _{-\infty}^{\infty} \frac
{dy}{2\pi} \, \phi ^{'}(x-y,1)
\, {\mbox {Im}} \ln \left [ 1+e^{iZ_3(y+i0)}\right ] + \nonumber \\
&+&\int _{-\infty}^{\infty} \frac {dy}{2\pi} \, \phi ^{'}(x-y,1/2)
\, Z_1(y) - 2 \int _{-\infty}^{\infty} \frac {dy}{2\pi} \, \phi
^{'}(x-y,1/2) \, {\mbox {Im}} \ln \left [ 1+e^{iZ_1(y+i0)}\right ]
\, . \nonumber
\end{eqnarray}
By symmetry considerations (note that $M_2=M_3=L/2$) we can infer
that $Z_2(x)=Z_3(x)$, so that we have to deal with only two
equations. We put again Fourier transforms into the game and
obtain
\begin{eqnarray}
\hat Z_1(k) &=& L \, \hat \phi (k,1/2)-\frac {1}{2\pi}\, \hat
{\phi ^{'}}(k,1) \, \hat Z_1(k) + 2 \, \frac {1}{2\pi}\, \hat
     {\phi ^{'}}(k,1) \, \hat L_1(k) + \nonumber \\
&+& 2 \, \frac {1}{2\pi} \, \hat {\phi ^{'}}(k,1/2) \, \hat Z_2(k)
- 4
 \, \frac {1}{2\pi} \, \hat {\phi ^{'}}(k,1/2) \, \hat L_2(k) \, , \nonumber \\
\hat Z_2(k) &=& -\frac {1}{2\pi} \, \hat {\phi ^{'}}(k,1) \, \hat
Z_2(k) + 2 \, \frac {1}{2\pi} \, \hat {\phi ^{'}}(k,1) \,
\hat L_2(k) + \nonumber  \\
&+&  \frac {1}{2\pi} \, \hat {\phi ^{'}}(k,1/2) \, \hat Z_1(k) - 2
\,  \frac {1}{2\pi} \, \hat {\phi ^{'}}(k,1/2) \, \hat L_1(k)
\nonumber
\end{eqnarray}
and, consequently,
\begin{eqnarray}
\hat Z_1(k)&=&\frac {L \, \hat \phi (k,1/2)}{1+\frac {1}{2\pi}\,
\hat
  {\phi ^{'}}(k,1)}+ 2 \frac {\frac {1}{2\pi}\, \hat {\phi ^{'}}(k,1)}
 {1+\frac {1}{2\pi}\, \hat {\phi ^{'}}(k,1)} \hat L_1(k) + \nonumber \\
&+& 2 \frac {\frac {1}{2\pi}\, \hat {\phi ^{'}}(k,1/2)} {1+\frac
  {1}{2\pi} \, \hat {\phi ^{'}}(k,1)} \hat Z_2(k) - 4
\frac {\frac {1}{2\pi}\, \hat {\phi ^{'}}(k,1/2)} {1+\frac
{1}{2\pi}\,
  \hat {\phi ^{'}}(k,1)}\hat L_2(k) \, , \nonumber  \\
\hat Z_2(k)&=& 2 \frac {\frac {1}{2\pi}\, \hat {\phi ^{'}}(k,1)}
{1+\frac {1}{2\pi}\, \hat {\phi ^{'}}(k,1)} \hat L_2(k) + \frac
{\frac {1}{2\pi}\, \hat {\phi ^{'}}(k,1/2)}
 {1+\frac {1}{2\pi}\, \hat {\phi ^{'}}(k,1)} \hat Z_1(k) - \nonumber  \\
&-& 2  \frac {\frac {1}{2\pi}\, \hat {\phi ^{'}}(k,1/2)} {1+\frac
{1}{2\pi}\, \hat {\phi ^{'}}(k,1)}\hat L_1(k) \, . \nonumber
\end{eqnarray}
No need to say that the usual short notation
\begin{equation}
\hat L_i(k)= \int _{-\infty}^{\infty} dx\ e^{-ikx}  {\mbox { Im}}
\ln \left [ 1+e^{iZ_i(x+i0)}\right ]  \, , \quad i=1,2 \, ,
\nonumber
\end{equation}
has appeared. Thus, clearly the $x$ Fourier transform of the
ubiquitous function $\phi(x,\xi)$,
\begin{equation}
\hat \phi (k,\xi)= -2\pi i e^{-\xi |k|} P\left (\frac {1}{k}
\right ) \, ,
\end{equation}
plays a central r\^ole to achieve
\begin{eqnarray}
\hat Z_1(k)&=&-L \, 2 \pi i \, \frac { e^{-\frac {|k|}{2}}}{1+e^{-|k|}} P \left  (\frac {1}{k} \right ) + 2 \frac { e^{-|k|}}{1+e^{-|k|}}\, \hat L_1(k) + \nonumber \\
&+& 2 \frac { e^{-\frac {|k|}{2}}}{1+e^{-|k|}} \, \hat Z_2(k) - 4
\frac { e^{-\frac {|k|}{2}}}{1+e^{-|k|}}\, \hat L_2(k) \, ,  \nonumber  \\
\hat Z_2(k)&=& 2 \frac { e^{-|k|}}{1+e^{-|k|}}\, \hat L_2(k) +
\frac { e^{-\frac {|k|}{2}}}{1+e^{-|k|}} \, \hat Z_1(k) - 2 \frac
{ e^{-\frac {|k|}{2}}}{1+e^{-|k|}} \, \hat L_1(k) \, .\nonumber
\end{eqnarray}
Eventually, these equations can be re-arranged in the clearer
manner
\begin{eqnarray}
\hat Z_1(k)&=&-L \, 2 \pi i\, e^{-\frac {|k|}{2}} \frac
{1+e^{-|k|}} {1+e^{-2|k|}} P \left  (\frac {1}{k} \right ) - 2
e^{-|k|} \frac {1-e^{-|k|}} {1+e^{-2|k|}} \, \hat L_1(k) - 4
\frac
{e^{-\frac {|k|}{2}}} {1+e^{-2|k|}} \, \hat L_2(k) \, , \nonumber \\
\hat Z_2(k)&=&-L \, 2 \pi i \, \frac {e^{-|k|}} {1+e^{-2|k|}} P
\left (\frac {1}{k} \right ) - 2 \frac {e^{-\frac {|k|}{2}}}
{1+e^{-2|k|}} \, \hat L_1(k) - 2  e^{-|k|} \frac  {1-e^{-|k|}}
{1+e^{-2|k|}} \, \hat L_2(k)\, , \nonumber
\end{eqnarray}
or, after anti-transforming, in the final NLIEs for the $Z$s
\begin{eqnarray}
Z_1(x)&=&F_1(x)+2 \int _{-\infty}^{\infty} dy \, G_{11}(x-y)
\,  {\mbox {Im}} \ln \left [ 1+e^{iZ_1(y+i0)}\right ] + \nonumber \\
&+& 2 \int _{-\infty}^{\infty} dy \, G_{12}(x-y)\,
 {\mbox {Im}} \ln \left [ 1+e^{iZ_2(y+i0)}\right ] \, , \label {nlin1} \\
Z_2(x)&=&F_2(x)+2 \int _{-\infty}^{\infty} dy \, G_{21}(x-y)\,
 {\mbox {Im}} \ln \left [ 1+e^{iZ_1(y+i0)}\right ] + \nonumber \\
&+& 2 \int _{-\infty}^{\infty} dy \, G_{22}(x-y)\,
 {\mbox {Im}} \ln \left [ 1+e^{iZ_2(y+i0)}\right ] \, . \label {nlin2}
\end{eqnarray}
The known functions in the previous equations are explicitly
\begin{eqnarray}
&&F_1(x)=2L \int _{0}^{\infty} dk \, \frac {\sin kx}{k} \, \frac
{\cosh \frac {k}{2}}{\cosh k} = 2L \, \arctan \left ( {\sqrt {2}}\sinh \frac {\pi x}{2} \right ) \, ,\nonumber \\
&&F_2(x)=L \int _{0}^{\infty} \frac {dk}{k} \, \frac {\sin
kx}{\cosh
  k} = L\, {\mbox {gd}} \frac {\pi x}{2} \, ,\nonumber \\
&&G_{11}(x)=G_{22}(x)=-\int _{0}^{\infty} \frac {dk}{2\pi}\, \cos
kx
\, \frac {1-e^{-k}}{\cosh k}  = \label {SO6functions} \\
&=& -\frac {1}{4}\frac {1}{\cosh \frac {\pi x}{2}}+\frac {1}{2 \pi
i}\, \frac {d}{dx} \ln \frac {\Gamma  \left ( 1+\frac
{ix}{4}\right )
\Gamma \left (\frac {1}{2}-\frac {ix}{4}\right )}{\Gamma \left (1-\frac {ix}{4}\right )\Gamma \left ( \frac {1}{2}+\frac {ix}{4}\right )} \, , \nonumber \\
&&G_{12}(x)=2\, G_{21}(x)=-\int _{0}^{\infty} \frac {dk}{\pi} \,
\cos kx
\, \frac {e^{\frac {k}{2}}}{\cosh k} \, = \nonumber \\
&=&-\frac {1}{\sqrt {2}}\frac {\cosh \frac {\pi x}{2}}{\cosh \pi
x} -\frac {1}{2}\frac {1}{\cosh \pi x} +\frac {1}{\pi i}\, \frac
{d}{dx} \ln \frac { \Gamma \left ( \frac {7}{8}+\frac {ix}{4}
\right )  \Gamma \left ( \frac {5}{8}-\frac {ix}{4} \right
)}{\Gamma \left ( \frac {7}{8}-\frac {ix}{4} \right ) \Gamma \left
( \frac {5}{8}+\frac {ix}{4} \right )} \, . \nonumber
\end{eqnarray}
Summarizing, the equations (\ref {nlin1}, \ref {nlin2}) are the
Non-Linear Integral Equations describing the anti-ferromagnetic
state (real solutions without holes to the Bethe equations) of the
$SO(6)$ symmetric chain (vector representation). This state
possesses zero momentum and the maximal energy. The latter will
receive an exact expression -- in terms of solutions of the NLIEs
(\ref {nlin1}, \ref {nlin2}) -- in the next subsection.

\subsection {The anomalous dimension}

The important result of \cite {MZ} is that the dilatation matrix
of scalar operators in ${\cal N}=4$ SYM at one loop can be mapped
to the hamiltonian of an integrable $SO(6)$ symmetric chain. In
terms of the Bethe roots, its eigenvalue $\gamma$ reads as
follows:
\begin{equation}
\gamma = \frac {\lambda }{16 \pi ^2} E\, , \quad
E=2\sum_{i=1}^{M_1} \frac {1}{u_{1,i}^2 + \frac {1}{4}} \, ,
\end{equation}
where $E$ is the chain energy. The maximal eigenvalue (anomalous
dimension) is obtained when considering the solution to the Bethe
equations containing real roots and no holes. For this
configuration, by the same arguments used in the previous
sections, a sum over the set $1$ of Bethe roots can be expressed
in terms of integrals involving $Z_1$:
\begin{eqnarray}
\sum _{i=1}^{M_1}  f(u_{1,i})&=&-\int _{-\infty}^{\infty} \frac
 {dx}{2\pi}\, f^{'}(x) \, Z_1(x) +2 \int _{-\infty}^{\infty}
\frac {dx}{2\pi} \, f^{'}(x) \, {\mbox {Im}} \ln \left [ 1+e^{iZ_1(x+i0)}\right ] = \nonumber \\
&=&  -\int _{-\infty}^{\infty} \frac {dk}{(2\pi)^2} \, \hat
{f^{'}}(k) \, \hat Z_1(-k) +2 \int _{-\infty}^{\infty} \frac
{dk}{(2\pi)^2} \, \hat {f^{'}}(k) \, \hat L_1(-k) \, . \nonumber
\end{eqnarray}
Inserting now the NLIE (\ref {nlin1}) for $Z_1$ yields
\begin{eqnarray}
&&\sum _{i=1}^{M_1}  f(u_{1,i})=L\int _{-\infty}^{\infty} \frac
{dk}{2\pi i}\, \hat {f^{'}}(k) \, e^{-\frac {|k|}{2}} \frac
{1+e^{-|k|}}
{1+e^{-2|k|}} \, P \left  (\frac {1}{k} \right )+ \nonumber \\
&+& \int _{-\infty}^{\infty} \frac {dk}{2\pi ^2}\, \hat {f^{'}}(k)
\, \frac {1+e^{-|k|}} {1+e^{-2|k|}} \, \hat L_1(-k) + \int
_{-\infty}^{\infty} \frac {dk}{4\pi ^2}\, \hat {f^{'}}(k) \, 4 \,
\frac {e^{-\frac {|k|}{2}}} {1+e^{-2|k|}} \, \hat L_2(-k) \, .
\nonumber
\end{eqnarray}
Upon specializing $f$ to be the single particle energy
\begin{equation}
E=2\sum_{i=1}^{M_1} \frac {1}{u_{1,i}^2 + \frac {1}{4}}
\Rightarrow f(u)=e(u)=\frac {2}{u^2 +\frac {1}{4}} \, , \quad \hat
{e^{'}}(k) = 4\pi ik \, {e^{-\frac {|k|}{2}}} \, ,
\end{equation}
we obtain
\begin{eqnarray}
E &=& 4L \int _{0}^{\infty} dk\, e^{-k} \frac
{1+e^{-k}}{1+e^{-2k}} + 2\frac {i }{\pi } \int _{-\infty}^{\infty}
dk \, k \,   \frac {\cosh \frac {k}{2}}
     {\cosh k} \, \hat L_1(-k) + \nonumber \\
&+& 2\frac {i }{ \pi }\int _{-\infty}^{\infty} dk \, k \, \frac
{1} {\cosh k } \hat L_2(-k) \, . \label {gamma}
\end{eqnarray}
The first term in the r.h.s. of  (\ref {gamma}) gives the leading
contribution when $L \rightarrow \infty$. It may be easily
evaluated
\begin{equation}
4L \int _{0}^{\infty} dk \, e^{-k} \frac {1+e^{-k}}{1+e^{-2k}} =4L
\int _{0}^{1} dx \, \frac {1+x}{1+x^2} = 2L \left ( \frac
{\pi}{2}+\ln 2 \right ) \, . \label {gammalead}
\end{equation}
We remark that (\ref {gammalead}) agrees with (5.10) of \cite
{MZ}. Hence, the maximal energy may be re-written as
\begin{eqnarray}
E&=&2L\left (\frac {\pi}{2}+\ln 2 \right )+ \int
_{-\infty}^{\infty} dx \, E_{1}(x)  \, {\mbox {Im}}
\ln \left [ 1+e^{iZ_1(x+i0)}\right ] +  \nonumber \\
&+&\int _{-\infty}^{\infty} dx \, E_{2}(x) \, {\mbox {Im}} \ln
\left [ 1+e^{iZ_2(x+i0)}\right ]  \, ,\label{enerso6}
\end{eqnarray}
where the last two terms contain the functions
\begin{equation}
E_{1}(x)=2{\sqrt {2}} \, \frac {d}{dx} \frac {\cosh \frac {\pi
x}{2}}{\cosh \pi x} \, , \quad E_{2}(x)=- \pi\, \frac {\sinh \frac
{\pi x}{2}}{\cosh ^2 \frac {\pi x}{2}} \, . \label {E(x)}
\end{equation}
Formula (\ref {enerso6}) is an exact expression for the energy in
terms of the solution of the NLIEs (\ref {nlin1}, \ref {nlin2}).
When $L\rightarrow \infty$, the last two terms provide the ${\cal
O}(1/L)$ corrections to the anomalous dimension. Analytical
computations up to the order $1/L$ will be the topic of the next
section.

\section{Analytic calculation of sub-leading order}
\setcounter{equation}{0}

It turns out that in all the discussed cases it is even possible
to single out the explicit sub-leading contribution to the energy
as $L\rightarrow \infty$. In fact, it is of order $1/L$ and comes
out in a rather standard way by following the strategy of the
`derivative lemma' \cite{DDV}.

Also, we need to mention that higher order corrections might be
extracted explicitly in the framework of NLIEs; for Kl\"umper was
able to compute the first logarithmic corrections (to the $1/L$
term) in the spin $1/2$-XXX Heisenberg chain starting by a NLIE
similar to ours, although using some numerical insights
\cite{klump} (see also \cite{affleck} and references therein for a
comparison with computations in a field theoretic framework). For
the time being, we are not interested in discussing the analytic
derivation of these contributions, but will return to them in a
subsequent paper \cite{35}. Nonetheless, their presence too is
expected in the multi-loop $SU(2)$ and $SO(6)$ cases, as motivated
in Section 6, where we will present our numerical results.

~~

{\bf XXX Heisenberg chain.} \ Let us first consider the Heisenberg
chain and the excitations over the anti-ferromagnetic state
described by holes. The contributions we want to evaluate are
contained in the integration term of (\ref{gammasu2}):
\begin{gather}
\Delta E (L)= \int _{-\infty}^{\infty }dy \, (-\pi)\, \frac {\sinh
\pi y}{\cosh^2 {\pi y}}
\, {\mbox {Im}}\ln \left[1+(-1)^{\delta}e^{iZ(y+i0)}\right]  = \notag \\
=\int _{-\infty}^{0}dy \, (-\pi)\, \frac {\sinh \pi y}{\cosh^2
{\pi y}}\,
{\mbox {Im}}\ln \left[1+(-1)^{\delta}e^{iZ(y+i0)}\right]+\label{asym} \\
+\int _{0}^{\infty }dy \, (-\pi)\, \frac {\sinh \pi y}{\cosh^2
{\pi y}} \, {\mbox {Im}}\ln
\left[1+(-1)^{\delta}e^{iZ(y+i0)}\right] \, . \notag
\end{gather}
In order to single out the order $1/L$ contributions, we perform
different changes of variables in each of the integrals in (\ref
{asym}), \be y=\theta - \frac {\ln 2L}{\pi} \ , \quad {\mbox
{for}}\  y<0 \ , \quad y=\theta + \frac {\ln 2L}{\pi} \  , \quad
{\mbox {for}}\ y>0 \ , \nonumber \ee and then we let $L\rightarrow
\infty$. In this limit we have
\begin{eqnarray}
\Delta E (L)&=& \frac{\pi}{L} \int _{-\infty}^{\infty }d\theta
\left\{ e^{\pi \theta}\ {\mbox {Im}}\ln \left[1+e^{iZ^{-}(\theta
+i0)}\right] - e^{-\pi \theta}{\mbox { Im}}\ln
\left[1+e^{iZ^{+}(\theta+i0)}\right]
 \right\}+ \nonumber \\
&+& o(1/L)\,, \label {DeltaE}
\end{eqnarray}
where the usual symbol $o(1/L)$ is used to indicate terms that
vanish faster than $1/L$ \be\label{varepsil} \lim _{L\rightarrow
\infty} o(1/L) \, L =0 \,. \ee The new functions \be
Z^{\mp}(\theta)=\lim _{L\rightarrow \infty} \left [ Z\left (\theta
\mp \frac {\ln 2L}{\pi} \right ) \pm \frac {\pi}{2}(H+L) \right ]
\, , \nonumber \ee satisfy the {\it kink} NLIEs:
\begin{eqnarray}
Z^{-}(\theta)&=&e^{\pi \theta} +  2 \int _{-\infty}^{\infty
}d\theta ^{\prime} \ G(\theta-\theta ^{\prime})\
{\mbox {Im}}\ln \left[1+ e^{iZ^{-}(\theta ^{\prime}+i0)}\right]\,, \nonumber \\
\label {kink1} \\
Z^{+}(\theta)&=&- e^{-\pi \theta} + 2 \int _{-\infty}^{\infty
}d\theta ^{\prime} \ G(\theta-\theta ^{\prime})\ {\mbox {Im}}\ln
\left[1+ e^{iZ^{+}(\theta ^{\prime}+i0)}\right]\,. \nonumber
\end{eqnarray}
We notice that the holes contribution in the NLIE (\ref {nlin22})
has become a constant and has been reabsorbed in a redefinition of
the counting function.

In summary, the integral term in (\ref {DeltaE}) gives the
contribution of order $1/L$ to the energy (as $L\rightarrow
\infty$). This term can be exactly computed by using the
derivative lemma based on the dilogarithmic function (for an
enunciation of this lemma see for instance Section 7 of \cite
{DDV}). The consequent result for $\Delta E(L)$ is \be \Delta E
(L) = \frac{\pi^2}{6L}  + o(1/L) \ee and it does not depend on the
holes, as long as their position remains finite for large $L$.

We can now evaluate the holes contribution to the energy
(\ref{gammasu2}). From (\ref{qholes}) we deduce the leading
behaviour \be x_h\sim \frac{2I_h+\delta-1}{L} \, , \ee namely
holes accumulate towards the point $x=0$. This leads to the
following contribution, \be - \sum_{h=1}^H \frac{\pi}{\cosh \pi
x_h} = -H\, \pi +\frac{\pi^3}{2L^2} \sum_{h=1}^H
(2I_h+\delta-1)^2+{\cal O}(1/L^3) \, , \ee from which we conclude
that hole excitations over the anti-ferromagnetic state do not
contribute to $1/L$ terms. In summary, the energy in given by \be
E=2L \ln 2 - \pi H +  \frac{\pi^2}{6L} + o(1/L)
 \, , \label {ESU2}
\ee where the symbol $o(1/L)$ is defined in (\ref {varepsil}).

~~

{\bf The $SU(2)$ multi-loop chain.}  In this case we will evaluate
explicitly the energy (\ref {energ4}) up to the order $1/L$ (in
the limit $L\rightarrow \infty$). Let us start by the term, \be
\Delta E(L,\lambda)=\int _{-\infty}^{\infty }dy\ e_1(y,\lambda)
{\mbox { Im}}\ln \left
[1+(-1)^{\delta}e^{iZ(y+i0,\lambda)}\right]\label{1energ2} \, ,
\ee where $e_1(x,\lambda)$ has been defined in (\ref{e1}). As
before, we split the integral (\ref {1energ2}) into two parts,
\begin{eqnarray}
\Delta E(L,\lambda)&=&\int _{-\infty}^{0}dy\ e_1(y,\lambda) {\mbox
{ Im}}\ln \left
[1+(-1)^{\delta}e^{iZ(y+i0,\lambda)}\right]+ \nonumber \\
&+&\int _{0}^{\infty}dy\ e_1(y,\lambda) {\mbox { Im}}\ln \left
[1+(-1)^{\delta}e^{iZ(y+i0,\lambda)}\right] \, , \nonumber
\end{eqnarray}
we perform the change of variables, \be y=\theta - \frac {\ln
2L}{\pi} \ , \quad {\mbox {for}}\  y<0 \ , \quad y=\theta + \frac
{\ln 2L}{\pi} \  , \quad {\mbox {for}}\ y>0 \ , \label {shift} \ee
and then we let $L\rightarrow \infty$. In this limit the two
integrals can be computed by using the residue method. The first
(second) one is evaluated after choosing a contour closing in the
lower (upper) complex $y$-plane. The poles of the integrand lie on
the imaginary axis, $k=\pm i \pi (2n+1) \, , \  n \geq 0$ and give
contributions proportional to $L^{-2n-1}$. Restricting to the
leading $1/L$ contribution in the limit $L\rightarrow \infty$, we
can write
\begin{eqnarray}
&& \Delta E(L,\lambda)=\frac {4\pi i}{L {\sqrt {\lambda}}}
J_1\left ( -i \frac {\sqrt {\lambda}}{2} \right ) \int
_{-\infty}^{\infty }d\theta \, e^{\pi \theta} \,
{\mbox {Im}}\ln \left[1+e^{iZ^{-}(\theta +i0,\lambda)}\right] - \nonumber \\
\label {Deltaloop} \\
&-&\frac {4\pi i}{L {\sqrt {\lambda}}}  J_1\left ( -i \frac {\sqrt
{\lambda}}{2} \right ) \int _{-\infty}^{\infty }d\theta \, e^{-\pi
\theta} \, {\mbox {Im}}\ln \left[1+e^{iZ^{+}(\theta
+i0,\lambda)}\right] +o(1/L) \,, \nonumber
\end{eqnarray}
where $o(1/L)$ satisfies (\ref{varepsil}) and, as before,
we have defined the new functions \be Z^{\mp}(\theta,\lambda)=\lim
_{L\rightarrow \infty} \left [ Z\left (\theta \mp \frac {\ln
2L}{\pi},\lambda \right) \pm \frac {\pi}{2}(H+L) \right ] \, .
\label {Zdef} \ee The equations satisfied by
$Z^{\mp}(\theta,\lambda)$ are obtained starting from (\ref
{nlieloops}), performing the shifts appearing in their definition
(\ref {Zdef}), then evaluating the leading contribution of holes
and the forcing term when $L\rightarrow \infty$. The holes term
gives, as before, $\mp \pi H/2$. On the other hand, the forcing term
can be evaluated by using the residue technique in a fashion
similar to the previous energy kernel calculation. Its
contribution is \be \mp \frac {\pi L}{2}\pm J_0\left ( i \frac
{\sqrt {\lambda}}{2} \right ) e^{\pm \pi \theta} + {\cal O}(1/L)
\, , \ee the first term coming from the residue in $k=0$, the
second from the residues in $k=\pm i\pi$ of the integrand of (\ref
{forza}). It follows that the equations satisfied by
$Z^{\mp}(\theta,\lambda)$ take the form
\begin{eqnarray}
Z^{-}(\theta,\lambda)&=& J_0\left ( i \frac {\sqrt {\lambda}}{2}
\right ) e^{\pi \theta} + 2 \int _{-\infty}^{\infty }d\theta
^{\prime} \ G(\theta-\theta ^{\prime})\
{\mbox {Im}}\ln \left[1+ e^{iZ^{-}(\theta ^{\prime}+i0,\lambda)}\right]\,, \nonumber \\
\label {kinkloop} \\
Z^{+}(\theta,\lambda)&=&-  J_0\left ( i \frac {\sqrt {\lambda}}{2}
\right )e^{-\pi \theta} +2\int _{-\infty}^{\infty }d\theta
^{\prime} \ G(\theta-\theta ^{\prime})\ {\mbox {Im}}\ln \left[1+
e^{iZ^{+}(\theta ^{\prime}+i0,\lambda)}\right]\,. \nonumber
\end{eqnarray}
We remark that we are still able to apply the derivative lemma in
order to compute the $1/L$ contributions of (\ref {Deltaloop}),
the only difference with the Heisenberg chain being that now the
functions involved are dressed with Bessel functions. The result
is \be \Delta E(L,\lambda)=\frac {4i}{L {\sqrt {\lambda}}}\frac {
J_1\left ( -i \frac {\sqrt {\lambda}}{2} \right )}{ J_0\left ( i
\frac {\sqrt {\lambda}}{2} \right )}
         \frac {\pi ^2}{6} + o(1/L) \,
       . \label{Deltaloop1}
\ee
We now consider the holes contribution in (\ref {energ4}): \be
-\sum_{h=1}^H \int_{-\infty}^{\infty} dk\ 2\pi\, e^{ikx_h}\,
\frac{\bessel{1}} {\sqrt{\lambda}\ k \cosh \frac{k}2} \, . \ee
From (\ref{qu_holes}) we deduce again the leading behaviour \be
x_h\sim \frac{2I_h+\delta-1}{L} \, . \ee It follows that the holes
contribution to the energy is \be -H \int_{-\infty}^{\infty} dk\
2\pi\, \frac{\bessel{1}} {\sqrt{\lambda}\ k \cosh \frac{k}2} +
{\cal O}(1/L^2)= -H \pi \sum _{l=0}^{\infty} \frac {(-\lambda
)^l}{2^{4l}l!(l+1)!}|E_{2l}| + {\cal O}(1/L^2) \, , \label
{hloops} \ee where $E_{2l}$ are the Euler numbers.

Therefore, summing up (\ref {Deltaloop1}, \ref {hloops}) with the
leading contribution to (\ref {energ4}) proportional to $L$, we
get that the energy of the multi-loop chain in the limit
$L\rightarrow \infty$ behaves as follows
\begin{eqnarray}
E&=&\frac{8\pi}{\sqrt{\lambda}}L \int _{0}^{\infty} dk\
\frac{\bessel{0}\bessel{1}}{k(e^k+1)}
 -H \pi \sum _{l=0}^{\infty}
\frac {(-\lambda )^l}{2^{4l}l!(l+1)!}|E_{2l}| + \nonumber
 \label{enloops} \\
&+&\frac {4i}{L {\sqrt {\lambda}}}\frac  { J_1\left ( -i \frac
{\sqrt {\lambda}}{2} \right )}{ J_0\left ( i \frac {\sqrt
{\lambda}}{2} \right )}
        \frac {\pi ^2}{6} + o(1/L) \,.
     \nonumber
\end{eqnarray}

~~

{\bf The $SO(6)$ symmetric chain.}  In an analogous way, one can
estimate exactly the coefficient of the $1/L$ correction to the
energy (\ref {enerso6}) of the $SO(6)$ chain (as $L\rightarrow
\infty$). This correction is contained in the two integrals of
(\ref {enerso6})
\begin{gather}
\Delta E (L)= \sum _{i=1}^2 \int _{-\infty}^{\infty }dy \, E_i(y)
\,
{\mbox {Im}}\ln \left[1+e^{iZ_{i}(y+i0)}\right]  = \notag \\
=\sum _{i=1}^2 \int _{-\infty}^{0}dy \, E_i(y) \,
{\mbox {Im}}\ln \left[1+e^{iZ_{i}(y+i0)}\right]+\label{asym2} \\
+\sum _{i=1}^2 \int _{0}^{\infty }dy  \, E_i(y) \, {\mbox {Im}}\ln
\left[1+e^{iZ_{i}(y+i0)}\right] \, . \notag
\end{gather}
As in the $SU(2)$ case, we perform different changes of variables
according to the region of integration, \be y=\theta - \frac
{2}{\pi} \ln 2L  \ , \quad {\mbox {for}}\  y<0 \ , \quad y=\theta
+ \frac {2}{\pi} \ln 2L  \  , \quad {\mbox {for}}\ y>0 \  ,
\nonumber \ee and then we let $L\rightarrow \infty$. In this limit
we get
\begin{gather}
\Delta E (L) = \label {DeltaE1}\\
= \frac{\pi}{L} \int _{-\infty}^{\infty }d\theta \left\{ \frac
{1}{\sqrt {2}}\, e^{\frac {\pi \theta}{2}}\ {\mbox {Im}}\ln
\left[1+e^{iZ^{-}_1(\theta +i0)}\right]+ e^{\frac {\pi
\theta}{2}}{\mbox { Im}}\ln \left[1+e^{iZ^{-}_{2}
(\theta+i0)}\right]
 \right\} + \nonumber \\
+ \frac{\pi}{L} \int _{-\infty}^{\infty }d\theta \left\{- \frac
{1}{\sqrt {2}}\, e^{-\frac {\pi \theta}{2}}\ {\mbox {Im}}\ln
\left[1+e^{iZ^{+}_1(\theta +i0)}\right]- e^{-\frac {\pi
\theta}{2}}{\mbox { Im}}\ln \left[1+e^{iZ^{+}_{2}
(\theta+i0)}\right]  \nonumber
 \right\} + \\
+ o(1/L) \,, \nonumber
\end{gather}
where $o(1/L)$ indicates terms that vanish as
(\ref{varepsil}). The new functions \be Z_1^{\mp}(\theta)=\lim
_{L\rightarrow \infty}\left[ Z_1\left (\theta \mp \frac {2}{\pi}
\ln 2L \right ) \pm L \pi \right ] \, , \quad
Z_2^{\mp}(\theta)=\lim _{L\rightarrow \infty} \left [ Z_2\left
(\theta \mp \frac {2}{\pi} \ln 2L \right ) \pm L \frac {\pi}{2}
\right ] \, , \nonumber \ee satisfy the {\it kink} NLIEs:
\begin{eqnarray}
Z_1^{\mp}(\theta)&=&\pm {\sqrt {2}}e^{\pm \frac {\pi \theta}{2}} +
2 \int _{-\infty}^{\infty }d\theta ^{\prime} \
G_{11}(\theta-\theta ^{\prime})\ {\mbox {Im}}\ln \left[1+
e^{iZ_1^{\mp}(\theta ^{\prime}+i0)}\right]+
\nonumber \\
&+& 2 \int _{-\infty}^{\infty }d\theta ^{\prime} \
G_{12}(\theta-\theta ^{\prime})\ {\mbox {Im}}\ln \left[1+
e^{iZ_2^{\mp}(\theta ^{\prime}+i0)}\right]
\,, \nonumber \\
\label {kink2} \\
Z_2^{\mp}(\theta)&=&\pm e^{\pm \frac {\pi \theta}{2}} + 2 \int
_{-\infty}^{\infty }d\theta ^{\prime} \ G_{21}(\theta-\theta
^{\prime})\ {\mbox {Im}}\ln \left[1+ e^{iZ_1^{\mp}(\theta
^{\prime}+i0)}\right]+
\nonumber \\
&+& 2 \int _{-\infty}^{\infty }d\theta ^{\prime} \
G_{22}(\theta-\theta ^{\prime})\ {\mbox {Im}}\ln \left[1+
e^{iZ_2^{\mp}(\theta ^{\prime}+i0)}\right] \nonumber \, .
\end{eqnarray}
Now, it happens that the two contributions of order $1/L$ in (\ref
{DeltaE1}) can be exactly computed by generalizing the derivative
lemma to the $SO(6)$ case -- a case with two coupled NLIEs. These
two contributions are equal and their sum gives the order $1/L$
contribution to the energy: \be \Delta E(L)=\frac {\pi ^2}{2L} +
o(1/L) \, . \label {anenso6} \ee We conclude that in the limit $L\rightarrow
\infty$ the energy (\ref {enerso6}) of the anti-ferromagnetic
state of the $SO(6)$ chain is given by \be E=2L \left (\frac
{\pi}{2} +\ln 2 \right )+\frac {\pi ^2}{2L} + o(1/L) \, . \ee This
finite size correction induces to think of a $c=3$ two-dimensional
conformal field theory.

\section{Numerical analysis}
\setcounter{equation}{0}

In the previous section we have computed the explicit contribution
to the energy up to order $1/L$ for growing $L$. However, the NLIE
formulation of the Bethe Ansatz equations allows us to work out
interesting and precise numerical computations as well (cf.
\cite{FMQR, FRT} as first numerical results within this approach).
The latter can be used, for instance, to confirm and to improve
analytical results. In this spirit, we have performed numerical
calculations in order to study contributions to the energy of
orders equal to and smaller than $1/L$, when $L$ is very large.

We show here few examples of our numerical solutions of the
equations. The most natural method is to solve them by iterations
\cite{FRT}. Even if the obtained equations are correct for all
numbers of holes and lengths of the chain, they are particularly
effective if one considers states with a small number of holes in
a `Fermi-Dirac sea' of real roots.

~~

\begin{figure}[h]\hspace*{13mm}
\includegraphics[width=0.8\linewidth]{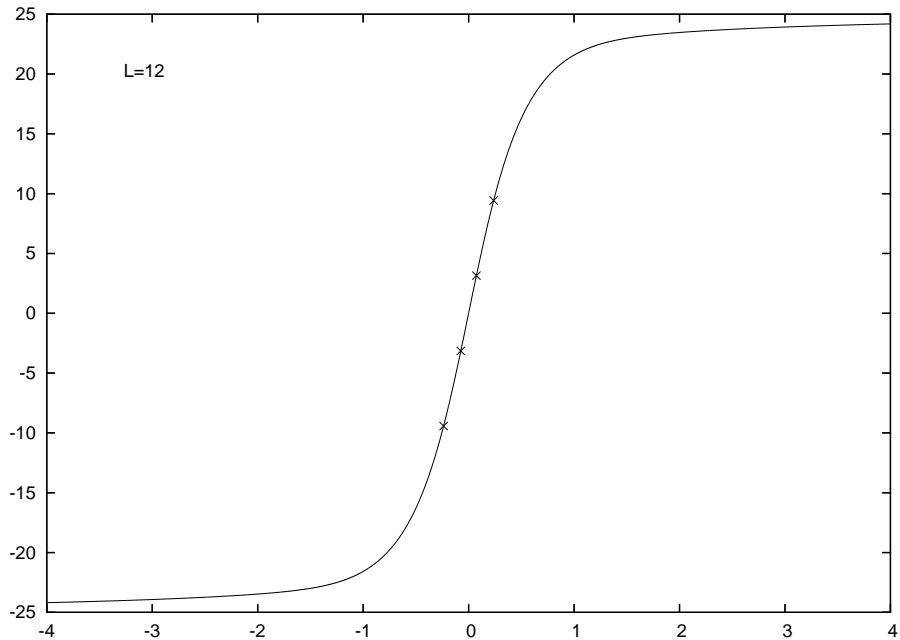}
\caption{\label{Zfig}Plot of the counting function $Z(x)$ versus
$x$ for the Heisenberg spin chain with $L=12$ sites. The position
of the four holes quantised by $I=(-1,\,0,\,1,\,2)$ is indicated
by the small crosses.}
\end{figure}

{\bf Heisenberg chain.}
\begin{figure}[h]\hspace*{13mm}
\includegraphics[width=0.8\linewidth]{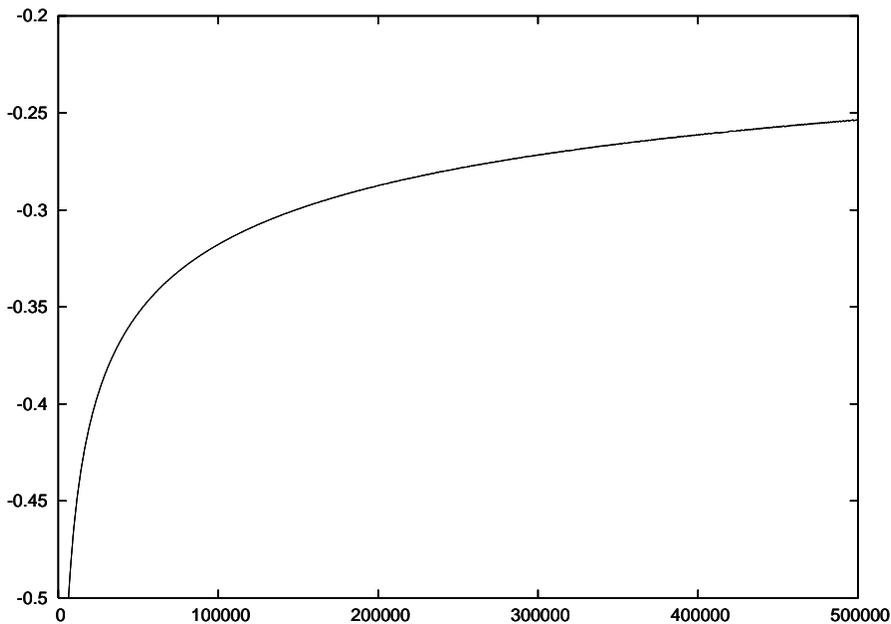}
\caption{Plot of $\Delta E(L)L-\frac{\pi^2}{6}$ versus $L$ for the
state $I=(-1,\,0,\,1,\,2)$ of the Heisenberg spin
chain.\label{4holes}}
\end{figure}
We consider a zero momentum state with four holes quantised by
$I=(-1,\,0,\,1,\,2)$ (see also Fig.~\ref{Zfig} for a prototypical
example of the behaviour of the counting function) and we follow
the evolution of $E$ with $L$. The goal is to show the order of
contribution of the various terms in (\ref{gammasu2}) when $L$ is
large. The leading contribution is explicit: $2L\ln 2$. From the
discussion in the previous section, we know that the holes
depending term gives a contribution $-H \pi + {\cal O}(1/L^2)$.
The contribution in $1/L$ is contained in the integral term
$(\ref{asym})$ that behaves as \be\label{deltae} \Delta E(L) =
\frac{\pi^2}{6L}+ o(1/L) \,  , \ee where $o(1/L)$ indicates
corrections that vanish faster than $1/L$. While trying to obtain
some insight on them we extended our analysis to chains of up to
five millions sites because we observed that the quantity $\Delta
E(L) \, L -\frac{\pi^2}{6}$ is extremely slow to vanish at growing
$L$, as clarified by Fig.~\ref{4holes}.

This behaviour would be perfectly consistent with the presence of
logarithmic terms \cite{affleck} (inside $o(1/L))$. Further, we expect
the first one of them to take the following form \be
\Delta E(L) \, L -\frac{\pi^2}{6} = \frac{c_1}{ \ln ^2 L} + \dots \ee
and we have found a good
agreement with the data coming from the numerical solution of the
NLIE. However, we refrain from giving any estimate of the constant
$c_1$, because numerical computations for very large chains
are technically difficult and are affected by growing numerical
errors. We expect to produce more precise data in the future to
better understand the additional terms in (\ref{deltae}).

The finite size corrections to the anti-ferromagnetic vacuum of
the Heisenberg chain have been extensively studied in condensed
matter literature \cite{affleck,klump}.
Taking into account the first logarithmic corrections we have \be
E = 2 L \, \ln 2 + \frac{\pi^2}{6 L} \left(1 + \frac{3}{8}
\frac{1} {\ln^3 L} +k_2 \frac{\ln \ln L}{\ln^4 L}
+\frac{k_3}{\ln^4 L}  \right) + \dots \, , \ee where the numerical
values of the constants $k_2$ and $k_3$ have been obtained by a
fit of the numerical data and have been found to be in agreement
with those of Karbach and M\"utter \cite{KM}. We remark that
logarithmic corrections to the energy of the anti-ferromagnetic
state are smaller than those for states containing holes. This
seems to be a general property of this state.

~~

{\bf $SU(2)$ at many-loops.} We can perform similar investigations
on $\Delta E(L,\lambda)$ in (\ref{1energ2}) when $L$ is large.
Even in this case, (with or without holes) we immediately see that
$o(1/L) \, L $ vanishes very slowly with $L$ and this suggests the
presence of logarithmic corrections. As pointed out in
\cite{klump} this logarithmic behaviour is related to the decay at
infinity of the kernel as a power. Since the multi-loop Bethe
equations of \cite{BDS} provide the same kernel as the Heisenberg
chain, the presence of such logarithmic corrections is somehow
expected.

We used numerical data, obtained for the same four hole state we
used in the Heisenberg case and for $\lambda=50$, to test the
agreement with the following guess for the logarithmic corrections
\be\label{espansione} \Delta E(L,\lambda)\, L - 0.7843670037...=
\frac{b_1}{ \ln ^2 L} +  \dots \, .\ee (on the left hand side we
have provided the numerical value of the coefficient of $1/L$ in
(\ref{Deltaloop1}) for $\lambda=50$). As in the Heisenberg case,
we refrain from giving any estimate of the constant $b_1$, because
of the growing numerical errors that affect computations for very
large chains and we postpone them to a forthcoming publication
\cite{35}.
\begin{figure}[h]\hspace*{13mm}
\includegraphics[width=0.8\linewidth]{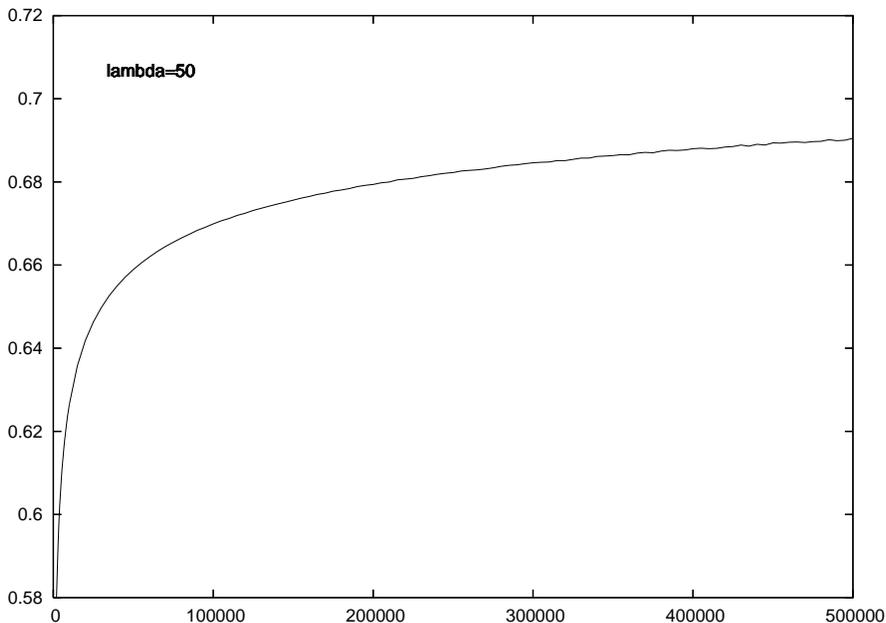}
\caption{\label{loopsDE}Plot of $\Delta E(L,\lambda) \, L$ versus
$L$ for the state $I=(-1\,,0\,,1\,,2)$ of the many-loops spin
chain with $\lambda=50$.}
\end{figure}

However, in close analogy with the Heisenberg chain, the proposed
functional form (\ref {espansione}) for the logarithmic
corrections was found to be in good agreement with the numerical
solution of the NLIE. In Fig.~\ref{loopsDE} it is possible to find
the plot of the data for $\Delta E(L,\lambda)\, L$ which have been
used for our analysis.

Finally, we guessed that the finite size corrections for $
E(L,\lambda) $ in the case of the anti-ferromagnetic state will
have the same structure of those corresponding to the Heisenberg
chain
\begin{eqnarray}
E(L,\lambda) =\frac{8\pi}{\sqrt{\lambda}}L \int _{0}^{\infty} dk\
\frac{\bessel{0}\bessel{1}}{k(e^k+1)}
 +\frac {4i}{L {\sqrt {\lambda}}}\frac  { J_1\left ( -i \frac
{\sqrt {\lambda}}{2} \right )}{ J_0\left ( i \frac {\sqrt
{\lambda}}{2} \right )}
        \frac {\pi ^2}{6} + o(1/L) \, ,
     \nonumber
\end{eqnarray}
with
\be o(1/L) = \frac{c(\lambda)}{L \ln^3 L} + \dots \ .  \ee
Again, in the case $\lambda = 50$, we obtained a good agreement
between the previous expression and the numerical data.

~~

\begin{figure}[h]\hspace*{13mm}
\includegraphics[width=0.8\linewidth]{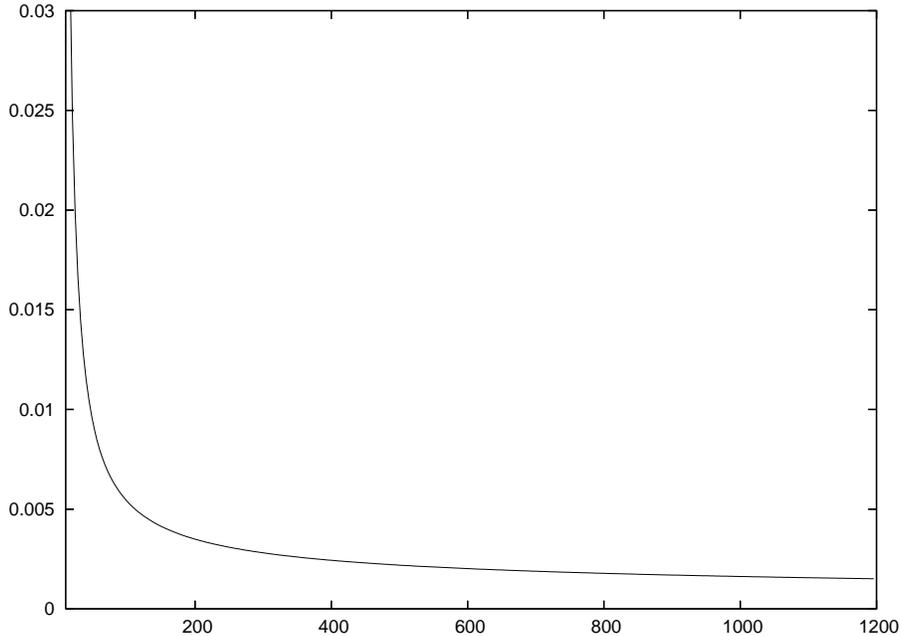}
\caption{\label{so6}Plot of $\Delta E(L) \, L-\frac{\pi^2}2$
versus $L$ for the anti-ferromagnetic state of the $SO(6)$ spin
chain.}
\end{figure}

{\bf The $SO(6)$ chain.} The same procedure used for the $SU(2)$
case applies to the $SO(6)$ chain, the main difference being that
now we need to solve two coupled equations. This does not imply
any additional difficulty in their solution, but simply increases
the computational time, and hence we prefer to limit our analysis
to the anti-ferromagnetic state for now. As before, we investigate
$\Delta E(L)\, L$ for large $L$ and compare numerical results
(coming from (\ref{enerso6})) with the analytical evaluation
(\ref{anenso6}), according to which \be \Delta E (L) \, L -
\frac{\pi^2}{2} = L \, o(1/L) \, . \ee Thanks to the various
insights from the above-discussed cases, we may for now limit our
analysis to chains with up to $L=1200$ sites: we intend to improve
the details in an incoming paper \cite{35} along with analytic
supports. Nevertheless, although not completely definitive, the
leading behaviour of $L \, o(1/L)$ seems to repeat that of the
Heisenberg ground state, namely $1/\ln^3L$. The corresponding
picture is in Fig.~\ref{so6}.

\section{Conclusive remarks}
\setcounter{equation}{0}

Briefly said, the novelty of this work may be summarised in the
derivation of the NLIEs and their application to computing anomalous
dimensions and momenta in ${\cal N}=4$ SYM theory when the number
of operators, $L$, is finite. More specifically, our attention has
been focussed on the $SU(2)$ scalar sub-sector for many loops and
the more general $SO(6)$ scalar sector for one loop. And for what
concerns the $SU(2)$ case we started by the conjectured Bethe
equations of \cite{BDS}, which are believed to be trustable at
least up to order $g^{2L-2}$. Actually, we have also re-derived,
following our own route, the know formul\ae \, for the isotropic
spin $1/2$ XXX Heisenberg chain as finite $L$ results for the one
loop correction in the $SU(2)$ scalar sub-sector: this effort was
at least useful to warm up and check the entire machinery.

These NLIEs, equipped with the quantisation conditions for the
holes, are totally equivalent to the Bethe equations we started
with. But of course they are much more effective for both the
analytic and numeric computations. For simplicity's sake, we
limited our analysis to the case of real roots, although the
introduced formalism can be easily extended to allow for the
possible presence of complex roots, as initiated by \cite{FMQR,
FRT}. Of course, because of its intrinsic nature this NLIE idea is
of easier applicability in presence of a very large number of real
roots, a limited number of holes and possibly of complex roots. In
terms of the spin chain this corresponds to focus the attention on
the anti-ferromagnetic state and excitations over it. However, the
anti-ferromagnetic configuration may receive particular interest
in SYM theory as that with the largest anomalous dimension; so, in
this respect it is antipodal to the ferromagnetic vacuum.

Nonetheless, the present formulation gives rise to compact and
exact expressions for the `observables' valid for any $L$. For
instance, the anomalous dimension assumes a form which, in
principle, could be exactly computed after solving the NLIE for
the counting function $Z(x)$ and after fixing the hole positions,
$x_h$, by the quantization conditions. In this respect, the
analytical results of Section 5 have given explicitly the
anomalous dimension up to the order $1/L$ (in the limit
$L\rightarrow \infty$). In addition, the numerical work of Section
6 has also shown the next-to-leading logarithmic dependence (on
$L$). In principle, the same ingredients might also be exploited
to obtain all the other conserved charges underlying integrability
\cite{FR1}. Therefore, a comparison with the string theory
integrability, first disclosed by \cite{BRP}, would be highly
instructive and desirable.

On the other hand, we would like to remark that the method used
here is quite flexible and can be applied to various integrable
models. Among them it is important to mention the Hubbard model
for its recent relevance in the computation of anomalous
dimensions of SYM \cite{RSS}. In this respect, it is not clear
whether the Hubbard chain might be the {\it ultimate} model in the
SYM/IM relation since there seems to be no trace of such a model
(or its symmetries) on the string side (not to say on the SYM
side).

Finally, it will be interesting to apply the present approach to
the case of large-$N$ QCD \cite{FHZ}, where the integrable
anti-ferromagnetic spin-1 XXX chain appears in the computation of
anomalous dimensions.

\section*{Acknowledgments}

We have the pleasure to acknowledge useful discussions with J.
Drummond, V. Fateev, L. Frappat and E. Ragoucy. We are all
indebted to EUCLID, the EC FP5 Network with contract number
HPRN-CT-2002-00325, which, in particular, supports the work of PG.
GF and MR want to thank APCTP, Pohang (South Korea) for kind
hospitality over the preliminary stages of this work. GF thanks
INFN for a post-doctoral fellowship and ISAS/SISSA, Trieste, with
a special acknowledgement to G. Mussardo's entourage. MR thanks
R. Sasaki and JSPS for the Invitation Fellowship for Research in Japan
(Long-term) L04716 and Lapth (Annecy), especially L. Frappat, for
kind invitation and support. DF thanks Leverhulme Trust for grant
F/00224/G, PRIN 2004 "Classical, quantum, stochastic systems with
an infinite number of degrees of freedom" for financial support
and the Theory Group in Bologna for a fantastic welcome.



\end{document}